\documentclass[aps,aps,10pt]{revtex4-2}

\usepackage{subcaption}
\usepackage{graphicx}
\usepackage{dcolumn}
\usepackage{bm}
\usepackage{setspace}
\usepackage{hyperref}
\usepackage{adjustbox}
\usepackage{array}
\usepackage[italicdiff]{physics}
\usepackage{float}
\usepackage[raggedrightboxes]{ragged2e}
\usepackage{amsmath} 
\usepackage{amssymb}
\usepackage{trimclip}
\usepackage{placeins}

\begin{document}
\def\AR{\clipbox{0pt 0pt .35em 0pt}{\textit{\bfseries A}}\kern-.05emR}

\textit{This paper is considering submission to Physics of Fluids.}
\title{\textbf{Modeling the effect of hydrodynamic wakes in dynamical models of large-scale fish schools}}

\author{Ji Zhou}
  \email{jzhou96@jhu.edu}
\author{Jung-Hee Seo}%
 \email{jhseo@jhu.edu}
\author{Rajat Mittal}%
 \email{mittal@jhu.edu}
\affiliation{%
 Mechanical Engineering, Johns Hopkins University\\
 Baltimore, MD 21218
}%

\date{\today}

\begin{abstract}
A novel model of the wake of swimming fish is developed and incorporated into a dynamical model of a fish school to explore the effect of hydrodynamics on the emergent behavior in schooling fish. The model incorporates well-established rules for attraction, alignment, and visual detection via a force-momentum balance in the surge, sway, and yaw directions, thereby allowing us to include the effects of body size, shape, and inertia in to the dynamics of fish motion. The key novelty of the model lies in the modeling of the hydrodynamics, which includes not only the potential flow induced by the body of the fish but also the vortex wakes generated by the fish. These hydrodynamic features, as well as the surge, sway, and yaw force coefficients, are parameterized via three-dimensional high-fidelity direct numerical simulations of a carangiform swimmer, thereby enabling a higher degree of realism in these models. The model is used to examine the effect of wake characteristics on the topology and movement of fish schools. The simulations indicate that these wake vortices lead to improved organization within the schools, especially in situations where the social forces are relatively weak.
\begin{description}
\item[keywords]
Fish Schools, Agent-Based Modeling, Collective Behavior, Social Interactions, Bio-Locomotion, Hydrodynamic Interactions.  
\end{description}
\end{abstract}
\maketitle

\doublespacing
\section{Introduction}
The collective motion of fish, commonly referred to as fish schooling, is a captivating natural phenomenon that has drawn significant attention in both experimental and computational studies \citep{partridge1982structure,pavlov2000patterns,zhou2021flow,weihs1973hydromechanics,zhou2022complex}. Schooling provides numerous benefits to fish \citep{pitcher1982fish,zhou2023effect,anras1997diel,zhou2024effect,mittal2023exploring}, including enhanced foraging efficiency, predator evasion, silent swimming, and reduced energetic costs, which are facilitated by sensory cues such as vision, pressure sensing, and lateral-line detection. While social interactions and sensory mechanisms are central to schooling dynamics, hydrodynamic effects also play a critical role in shaping these interactions and their outcomes \citep{filella2018model,maertens2017optimal,seo2022improved,mabrouk2024group}.

Hydrodynamic interactions primarily stem from the wakes generated by swimming fish, which can extend multiple body lengths downstream. These wakes induce forces that trailing fish may exploit to reduce drag, enhance thrust, and stabilize their motions \citep{pan2020computational,maertens2017optimal,ashraf2017simple,verma2018efficient,seo2022improved,wei2022passive,saadat2021hydrodynamic,han2024revealing}. At the same time, perturbed flow fields provide sensory cues that enable fish to position themselves optimally within a school, even in environments where visibility is limited \citep{ormonde2021two,seo2022improved,wei2022passive,mckee2020sensory}. However, directly quantifying the effects of hydrodynamics in fish schools remains a formidable challenge  \citep{timm2024multi,liao2007review}. Experimental methods often face significant difficulties in capturing wake dynamics and disentangling their influence from the inherent behavioral variability of fish. Additionally, high-fidelity computational approaches such as direct numerical simulations (DNS) are computationally expensive and impractical for studying large aggregates.

As a result, much of the existing research on schooling dynamics has relied on agent-based models, which are computationally efficient and focus on behavioral rules such as attraction, alignment, and avoidance (see Table \ref{table:previousModels}). While these models have successfully captured vivid emergent schooling patterns, they largely neglect hydrodynamic interactions. Recent efforts to incorporate hydrodynamics into agent-based frameworks, such as modeling potential flows around fish, have demonstrated that these effects can significantly alter schooling dynamics \citep{filella2018model,huang2024collective}. However, potential flow models fail to account for the persistent vortex wakes generated by swimming fish, which carry critical information about fish size, tail-beat frequency, and kinematics, influencing trailing fish over much longer distances \citep{viana2005potential,seo2022improved}.

This study introduces a novel agent-based model that addresses these limitations by incorporating a phenomenological representation of fish vortex wakes parameterized using DNS. Unlike previous models, our approach explicitly accounts for the hydrodynamic forces and moments induced by these wakes, integrating them into a framework governed by Newtonian mechanics. This model allows us to investigate how hydrodynamic effects shape the stability, formation topology, and energetic efficiency of fish schools, providing critical insights into the interplay between social behavior and fluid dynamics.

By bridging the gap between behavior-focused agent-based models and hydrodynamic realism, this study offers a timely contribution to the field. Given the increasing interest in understanding fish schooling from both biological and fluid dynamics perspectives, our work provides a valuable reference for researchers seeking to quantify the role of hydrodynamics. Furthermore, the presented approach complements experimental and DNS studies, offering a computationally efficient alternative for exploring large-scale schooling dynamics. We believe this framework will serve as a foundation for future investigations into the interplay between hydrodynamics and behavior in collective motion, advancing the field of bio-inspired fluid mechanics.

\renewcommand{\arraystretch}{1.5}
\begin{table}[h]
\caption{\label{tab:table1}Collection of several fish schooling models with key features.}
\scriptsize 
\begin{adjustbox}{width=\textwidth,center}

\begin{tabular}{lllllllll}
\toprule 
\textbf{Reference}                                                             & \textbf{Year} & \textbf{Dimension} & \textbf{Newton's 2nd Law} & \multicolumn{2}{l}{\textbf{Velocity}}    & \textbf{Sensing Range} & \textbf{Neighbor Scaling}    & \textbf{Hydrodynamics} \\ \hline
\begin{tabular}[c]{@{}l@{}}M{\"u}ller-Hartmann \\ \& Zittartz \citep{muller1974new}\end{tabular} & 1974          & 2D                 & Y                         & \multicolumn{2}{l}{Constant}             & Uniform                & Averaging                    & N                      \\
Huth \& Wissel \citep{huth1992simulation}                                                           & 1990          & 2D                 & N                         & \multicolumn{2}{l}{Gamma   Distribution} & Biased                 & Averaging   \& Bias in front & N                      \\
Sannomiya et al. \citep{sannomiya1990application}                                                      & 1990          & 2D                 & Y                         & \multicolumn{2}{l}{Equation   Governed}  & Uniform                & Averaging                    & Drag                   \\
Terzopoulos et al. \citep{terzopoulos1994artificial}                                                     & 1994          & 3D                 & Y                         & \multicolumn{2}{l}{Equation   Governed}  & Biased                 & Rules                        & Drag, Lift             \\
Czir{\'o}k et al. \citep{czirok1997spontaneously}                                                          & 1997          & 2D                 & N                         & \multicolumn{2}{l}{Constant}             & Uniform                & Averaging                    & N                      \\
Vab{\o} \& N{\o}ttestad \citep{vabo1997individual}                                                   & 1997          & 2D                 & N                         & \multicolumn{2}{l}{Equation   Governed}  & Uniform                & Averaging                    & N                      \\
Couzin et al. \citep{couzin2002collective}                                                         & 2002          & 3D                 & N                         & \multicolumn{2}{l}{Constant}             & Biased                 & Averaging                    & N                      \\
Takagi et al. \citep{takagi2004mathematical}                                                       & 2004          & 2D                 & Y                         & \multicolumn{2}{l}{Gamma   Distribution} & Uniform                & Averaging                    & N                      \\
Calovi et al. \citep{calovi2014swarming}                                                      & 2014          & 2D                 & N                         & \multicolumn{2}{l}{Equation   Governed}  & Biased                 & Bias   in front              & N                      \\
Filella et al. \citep{filella2018model}                                                      & 2018          & 2D                 & N                         & \multicolumn{2}{l}{Equation   Governed}                       & Biased        & Bias in front              & Potential flow \\
Wang et al. \citep{wang2023modeling}                                                      & 2023          & 2D                 & N                         & \multicolumn{2}{l}{Constant}                       & Uniform        & Averaging              & N\\
\toprule 
\end{tabular}
\end{adjustbox}
\label{table:previousModels}
\end{table}
\section{\label{sec:methods}Methodology}
\subsection{Dynamical Model of Schooling Fish}
\begin{figure}[h]
    \centering
    {\includegraphics[width=1\textwidth]{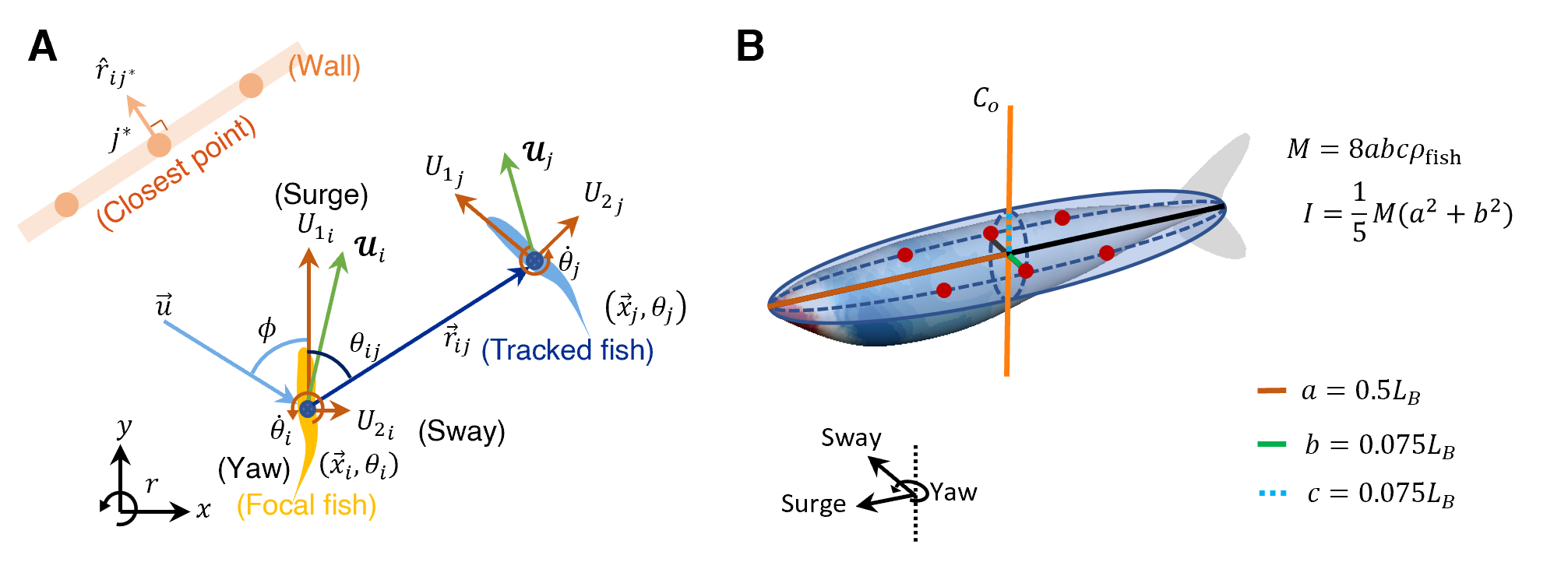}}
    \caption{The Lagrangian coordinate system and fish anatomy. \textbf{A}: Illustration of the interaction between fish $i$ (focal fish) and fish $j$ (tracked fish), from the perspective of fish $i$. $U_1$ and $U_2$ are velocities in the surge and sway directions, respectively. $\boldsymbol{\mathcal{U}}$ is the velocity vector of the fish. $\dot{\theta}$ represents the rotational velocity in the yaw direction. $(\vec{x},\theta)$ represents the location of fish in the Eulerian coordinate system and body orientation. $\vec{r}_{ij}$ is the direction vector from fish $i$ to fish $j$. $\theta_{ij}$ is the angle between $\vec{r}_{ij}$ and $U_1$. $\vec{u}$ is the relative velocity from the fluid to the fish body with the angle of attack, $\phi$. The closest point on the boundary wall is noted as $j^*$ and the unit vector normal to the wall is noted as $\hat{r}_{ij^*}$. \textbf{B}: Dimensions of the mackerels were normalized by the body length, $L_B$. The body was simplified as a 3D ellipsoid with the major axis, $a$, and minor axes $b=c$. The rotation centroid in the yaw direction is noted as $C_o$. The sensing nodes on the surface of the fish are marked as red dots.}
    \label{fig:fishAnatomy}
\end{figure}

In this study, we develop a dynamical model of fish schooling by coupling a Newtonian physics-based framework for fish movement with an ansatz for sensing and social interactions among fish. The model assumes that fish are 3D entities swimming within the same horizontal plane, which simplifies the problem to two dimensions (2D). This simplification is commonly used in schooling models, as the majority of schooling behaviors, such as alignment, attraction, and formation topology, occur predominantly in a horizontal plane and have been successfully captured by existing 2D models (Talbe ~\ref{table:previousModels}). To ensure accuracy, hydrodynamic forces and torques in this model are parameterized using 3D DNS of caudal fin swimmers, enabling the wake-induced dynamics to be represented with high fidelity. This combination of 2D modeling and 3D DNS parameterization allows the model to retain computational efficiency while accurately capturing the essential hydrodynamic interactions between fish. The governing equations for fish motion in the surge, sway, and yaw degrees of freedom are presented below (see Fig.~\ref{fig:fishAnatomy}):
\begin{align}
    \text{Surge:} \quad & M \pdv{U_1}{t} = P_1 + F_1 + F_s \label{eq:1} \\
    \text{Sway:} \quad & M \pdv{U_2}{t} = F_2 \label{eq:2} \\
    \text{Yaw:} \quad & I \pdv{\omega}{t} = T_3 + T_s + T_{\text{wall}} \label{eq:3}
\end{align}
Here, \( U_1 \), \( U_2 \), and \( \omega \) represent the surge, sway, and yaw (angular) velocities, respectively. The mass and moment of inertia of the fish are denoted by \( M \) and \( I \). The propulsive force generated by the caudal fin is \( P_1 \), while \( F_1 \), \( F_2 \), and \( T_3 \) represent the hydrodynamic forces and torque induced by flow interactions with the fish. These forces include the effects of hydrodynamic interactions between neighboring swimmers. The social interaction between fish is represented by \( F_s \) and \( T_s \), which model forces and torques due to alignment, attraction, and repulsion. Finally, \( T_\text{wall} \) is the torque that prevents fish from colliding with boundaries.

The wall-avoidance torque, \( T_\text{wall} \), is only modeled in the yaw direction. This simplification is based on observations that fish primarily rely on turning maneuvers to avoid collisions rather than active lateral movements, which are rare in natural schooling behavior. Such turning maneuvers align with well-documented escape responses, like C-starts, where yaw motions are sufficient for collision avoidance \citep{calovi2014swarming,witt2015hydrodynamics,wohl2007predictive}. Consequently, active lateral movement is excluded from this model.

The parameters in this model were chosen based on a combination of numerical simulations and observations of fish behavior. Hydrodynamic forces and torques (\( F_1, F_2, T_3 \)) are parameterized using results from our DNS, ensuring fidelity in capturing wake-induced effects. Social interaction parameters (\( F_s, T_s \)) are guided by observed schooling patterns and were further evaluated in this study to reveal the impact of social behavior on fish distributions in schooling.

Closure of the above equations requires information on fish anatomy, propulsive forces, hydrodynamic forces and torques, and social interactions. The fish are tracked as Lagrangian entities, while the flow induced by their motion is computed in a Eulerian framework. The equations of motion are integrated using a 4th-order Runge-Kutta scheme, with the modeling of induced velocities described in subsequent sections.
\subsection{\label{sec:fishAnatomy}Fish Anatomy}
The current fish model is based on a previously used model of a mackerel (\textit{Scomber scombrus})  \citep{seo2022improved}, which is a prototypical carangiform swimmer. For the purposes of the current model, the body of the fish is simplified into an ellipsoid with the shape parameters shown in Fig. \ref{fig:fishAnatomy}\textbf{B}. The frontal area, side area and volume are then obtained as $A_o=4bc$,  $A_s=4ac$ and $V_s=8(abc)$, respectively. Assuming a uniform tissue density equal to water, the centroid $(C_o)$, mass $(M)$ and moment-of-inertia $(I)$ are also easily obtained. All of these parameters can be adjusted as per the anatomy of the fish under consideration.
\subsection{\label{sec:propulsion}Propulsion}
The fish is assumed to propel itself by flapping its caudal fin, and we specify the steady swimming speed of the solitary fish as $U_o$. For steady swimming, the propulsion force $(P_1)$ balances the drag on the body of the fish, and for the high Reynolds numbers relevant to this situation, the drag is assumed to be given by $D_0=\tfrac{1}{2}{\rho}U_o^2L_B^2C_{Do}=P_1$, where $C_{Do}$ is the coefficient of drag for the fish swimming in a straight line. The flapping frequency $\mathcal{F}$ of the caudal fin that generates the propulsion is assumed to follow the empirical relationship given by \citep{bainbridge1958speed}, i.e. $\mathcal{F}=\tfrac{4}{3}(U_o/L_B +1)$. 
\subsection{\label{sec:hydrodynamicForcesAndMoment}Hydrodynamic Forces and Moment}
\begin{figure}[h]
    \centering
    {\includegraphics[width=1\textwidth]{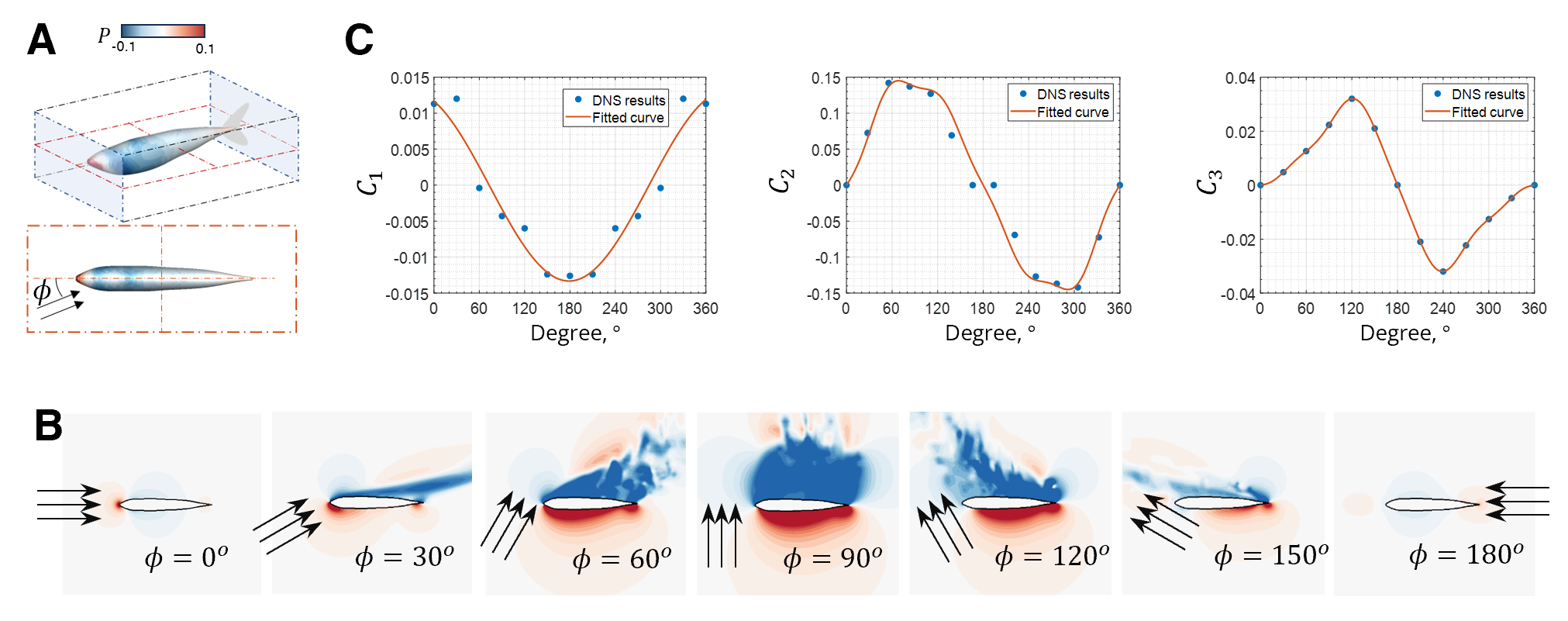}}
    \caption{Direct numerical simulations of a static fish under different angles of attack. \textbf{A}: Illustration of a stationary fish inside a 3D domain. For better illustration, the size of the box in \textbf{A} is not the actual domain size in simulations. \textbf{B}: Contour plots of the pressure under different angles of attack at the section marked in dashed red lines in \textbf{A}. $\phi$ is the angle between the attacking flow and the negative surge direction, ranging from 0 to 360 degrees. \textbf{C}: Fitted $C_1$, $C_2$, and $C_3$ using sinusoidal functions. Sample data were computed from DNS.}
    \label{fig:CdClCm}
\end{figure}

The terms $F_1$, $F_2$, and $T_3$ are the hydrodynamic force and torque induced by the flow relative to the fish and these are modeled as follows – considering the flow velocity relative to the fish has a magnitude $\mathcal{U}$ and is at an angle of $\phi$ from the surge direction, the hydrodynamic drag, lift and moment exerted on the fish can be expressed as:
\begin{equation}
F_1=\tfrac{1}{2}{\rho}\mathcal{U}^2L_B^2C_1; \quad
F_2=\tfrac{1}{2}{\rho}\mathcal{U}^2L_B^2C_2; \quad 
T_3=\tfrac{1}{2}{\rho}\mathcal{U}^2L_B^3C_3,
\label{F1F2Tw}
\end{equation}
 where $C_1$, $C_2$, and $C_3$ are the drag, lateral (sway), and yaw moment coefficients, respectively. Furthermore, ${\mathcal{U}}=(-U_1+u_1,-U_2+u_2)$ is the flow relative to the fish, where  $\vec{u}=(u_1,u_2 )$ is the velocity perturbation induced by the other fish in the school.
The coefficients $C_1$, $C_2$, and $C_3$ are functions of the angle $\phi$ of the flow relative to the body axis and these functional dependencies are determined via direct numerical simulations (\textrm{DNS}) of flow past a stationary fish at different angles-of-attack (Fig. \ref{fig:CdClCm}(\textbf{A, B})). Seven \textrm{DNS} were conducted using \emph{ViCar3D} \citep{mittal2008versatile}, an extensively used and validated \citep{solano2022perimeter,zhou2024effect,seo2022improved} immersed-boundary-based flow solver. Fig. \ref{fig:CdClCm}\textbf{C} shows the results of the simulations and the best fit periodic functions for each coefficient.

Given these coefficients, the angle of the flow relative to the fish (given by
$\phi=\textrm{tan}^{-1}[\frac{-U_1+u_1}{-U_2+u_2}]$) and the flow velocity $\mathcal{U}$ relative to the fish, the hydrodynamic forces and moment on the fish can be computed by the expressions given above. The location, size, and orientation of the body are represented by the six points shown in Fig. \ref{fig:fishAnatomy}\textbf{A}, and at each time step, we estimate $\mathcal{U}$ for the focal fish by averaging the velocity $\mathcal{U}$ on the six body-oriented points. With the resultant $\mathcal{U}$, we can compute the $\phi$, the corresponding $F_1$, $F_2$, and $T_\omega$ of the focal fish.

\subsection{Flow Velocity Induced by a Swimming Fish}
\begin{figure}[h]
    \centering
{\includegraphics[width=1\textwidth]{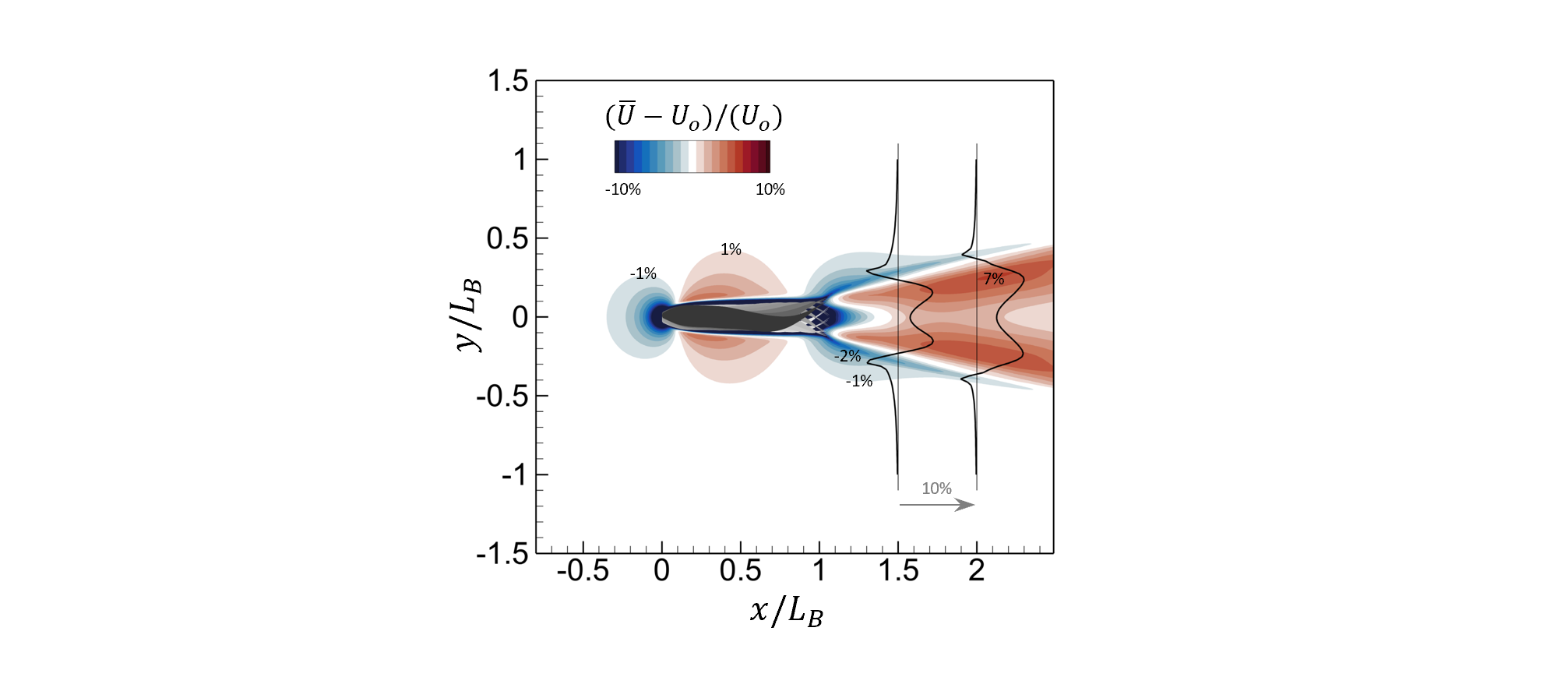}}
    \caption{Time-averaged contour plot of the induced velocity in the surge direction. Velocity profiles are plotted in black curves across the wake region.}
    \label{fig:dns}
\end{figure}

The final ingredient in the hydrodynamic interaction model is the determination of the velocity perturbation $\vec{u}=(u_1,u_2)$ induced by swimming fish. Some previous models of fish schooling have assumed that this perturbation is negligible (see, for instance, \citep{lopez2012behavioural}). However, a recent study (\citep{filella2018model}) introduced a dipole velocity perturbation associated with the potential flow induced by a moving point particle in 2D flow and found that the interaction effects introduced by this model induce complex patterns in the topology and movement of the fish school. This study indicated the possible importance of hydrodynamic interactions in fish schooling; however, studies suggest that the dominant feature in hydrodynamic interactions between fish is the vortex wake (\citep{seo2022improved,liao2007review}). Indeed, Fig. \ref{fig:dns} shows a time-averaged wake profile of a swimming fish from our previous \textrm{DNS} (\citep{seo2022improved}), which shows that while the flow perturbation associated with the fish body itself decays down to 1\% of the swimming velocity at a distance of a half a body-length, the perturbations induced by the flapping caudal fin are indeed stronger - up to 7\% of the surge velocity. These perturbations extend several body lengths into the wake. Our more recent studies indicate that the wake generated by the leading fish strongly correlates with the swimming efficiency \citep{seo2022improved} and acoustic footprints \citep{zhou2024effect}, further proving the significance of the wake. Given this dominance of the wake, we incorporate this feature alongside the potential flow induced by the body into our fish schooling model. 

\begin{figure}[h]
    \centering
    {\includegraphics[width=1\textwidth]{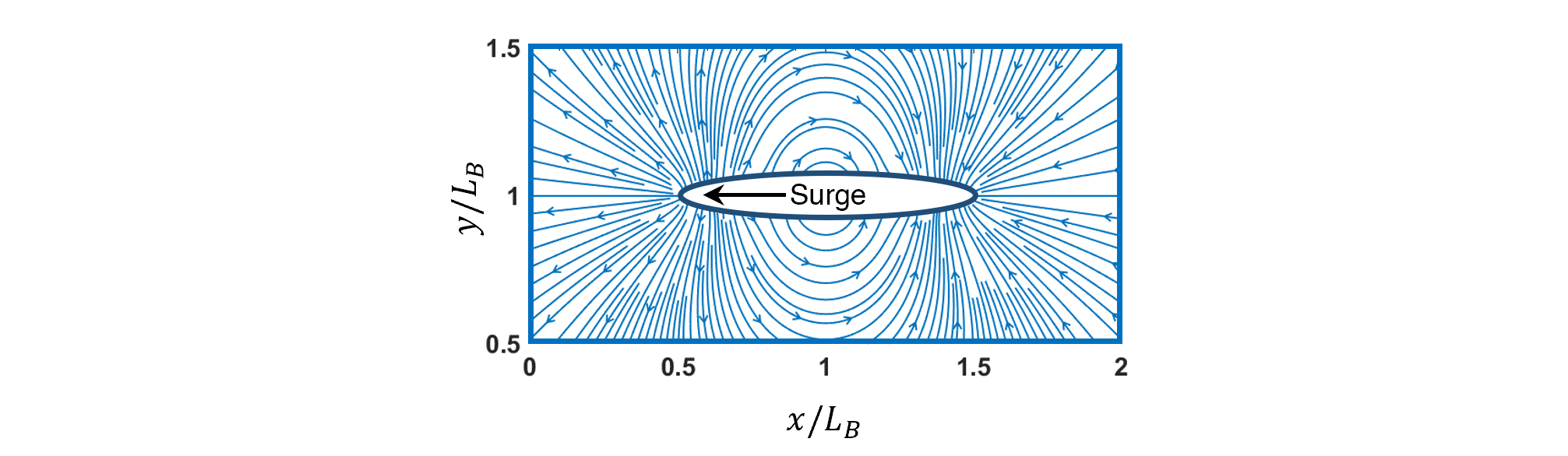}}
    \caption{Potential flow field of a single fish heading to the negative $x$ direction. The ellipsoid represents the fish body. The streamlines of the potential flow of a single swimming fish, mean flow subtracted.}
    \label{fig:streamline}
\end{figure}

The velocity perturbation induced by the fish is decomposed as $\vec{u}=\vec{u}_p+\vec{u}_w$, where $\vec{u}_p$ and $\vec{u}_w$ correspond to the potential flow due to the body and the wake flow, respectively, with the latter being primarily associated with the caudal fin motion. To estimate the potential flow due to the body of the fish, we approximate the fish body as an ellipsoid with the aspect ratio, $a:b:c = 1:0.15:0.15$. Four pairs of sink and source are located along the axis of the body, and the total body potential for this configuration is given as follows:
\begin{equation}
\phi=Ux-\frac{1}{4\pi}\sum_{m=1}^{4} \Phi_m[\frac{1}{\sqrt{(x+a_m)^2+y^2+z^2}}-\frac{1}{\sqrt{(x-a_m)^2+y^2+z^2}}],
\label{eq:4}
\end{equation}
where the position vector is relative to the location of the fish centroid. The locations $a_m$ and strengths $\Phi_m$ of these singular pairs are determined (Table. \ref{table:parametersofpotential}) via an iterative method to minimize the penetration velocity across the ellipsoid's surface. The resulting flow velocity is subsequently estimated via Eqn. \ref{eq:5}, and Fig. \ref{fig:streamline} shows the streamlines of the potential flow for a solitary swimming fish. The potential velocity ($u_p$,$v_p$) is given by
\begin{equation}
u_p=\pdv{\phi}{x};\quad v_p=\pdv{\phi}{y}. 
\label{eq:5}
\end{equation}

\begin{figure}[ht]
    \centering
    {\includegraphics[width=1\textwidth]{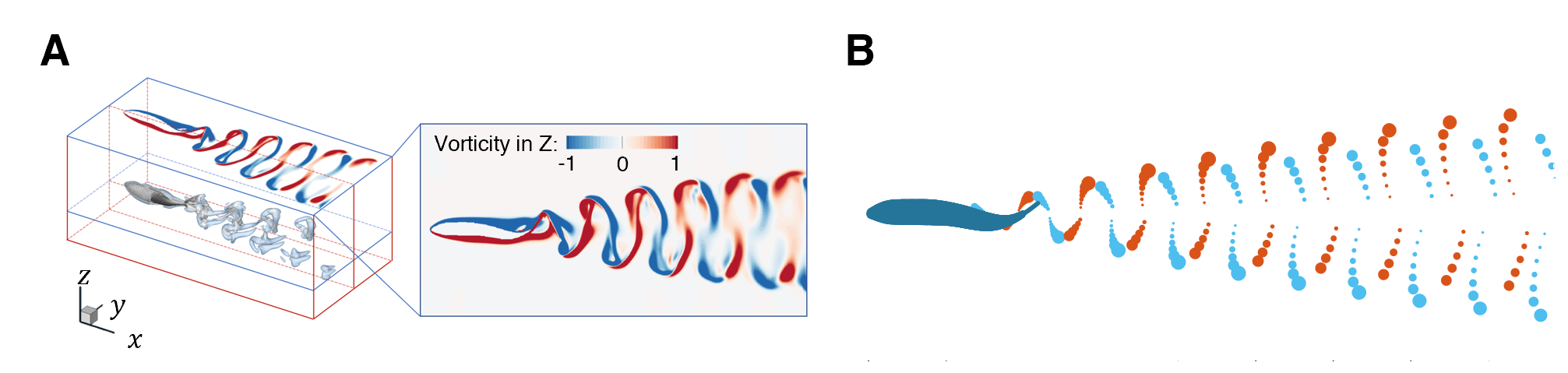}}
    \caption{Instantaneous snapshots of fish generated wake from the direct numerical simulation \textbf{A} and the presenting model \textbf{B}. \textbf{A}: The vorticity contour is from the cross-section of the DNS. \textbf{B}: Illustration of the reconstructed wake of a single fish. The relative size of the red and blue dots represents the relative circulation $\mathit{\Gamma}$ of the vortices.}
    \label{fig:wakeIllustrationDNS}
\end{figure}

The next step is to incorporate a model for the wake velocity perturbation $\vec{u}_w$, and we rely on our \textrm{DNS} results to inform this model as well. Fig. \ref{fig:wakeIllustrationDNS}\textbf{A} shows the vorticity field on the central plane generated by the carangiform kinematics of the fish. In Fig. \ref{fig:wakeIllustrationDNS}\textbf{B}, the vorticity field consists of a set of counter-rotating vortex pairs that are arranged in an oblique pattern \citep{dong2006wake,seo2022improved} along the wake. Note that this wake pattern is quite different from the inverse Karman vortex street observed in 2D configurations and is a more realistic representation of the vortex wake generated by a carangiform swimmer in reality. 

We note from the inset in Fig. \ref{fig:wakeIllustrationDNS}\textbf{A} that each half of the tail-beat generates an elongated vortex of alternating sign which is stronger at the two ends compared to the middle region. We synthesize a discrete vortex model here that mimics this phenomenology.  In the current model, each half tailbeat generates a set of Rankine vortices along a line segment with the same rotational direction but a non-uniform vortex strength $\mathit{\Gamma}$, and the rotation direction of these vortices alters in every half-beat of the tail. The tailbeat frequency can be computed according to the Bainbridge equation ($\mathcal{F}=\tfrac{4}{3}(U_o/L_B +1)$ \citep{bainbridge1958speed}) given the swimming velocity. Since this model is dynamic and the tailbeat frequency varies according to the decision of acceleration or deceleration, the real-time tailbeat frequency is coupled with the social force term $F_s$. The oblique angle of the wake is controlled by $V_s$, which, based on the \textrm{DNS} is 25\% of the fish's forward swimming velocity. The non-linear functions that describe the location and strength of these vortices are included in Table. \ref{table:parametersoffishwake}. The rate of decay of the circulation of these vortices is fit to exponential functions which are also parameterized from the \textrm{DNS} data (Table. \ref{table:parametersoffishwake}). The circumferential induced velocity at any point due to a given vortex is given by 
\begin{equation}
    \boldsymbol{u_w}=\frac{\mathit{\Gamma}}{2{\pi}r^{'}} \label{eq:u_w},
\end{equation}
where $r'$ is the radial distance of the point relative to the center of the vortex. The average induced velocity at the 6 points that define the body of the focal fish (see Fig. \ref{fig:fishAnatomy}\textbf{B})  is computed by summing up the induced velocity of all the vortices in the computational domain Eq. \ref{eq:sumu} (except for those vortices that are inside its own ellipsoidal body). Thus,
\begin{equation}
    \vec{u}_{\text{avg}} = \frac{1}{N_b} \sum_{i=1}^{N_b} \left( \vec{u}_{p,i} + \sum_{j=1}^{N_w} \vec{u}_{w,j,i} \right),
    \label{eq:sumu}
\end{equation}
where $N_b$ and $N_w$ are the number of points on fish body and the total number of vortices in the domain, respectively.

\begin{figure}[ht]
    \centering
    {\includegraphics[width=1\textwidth]{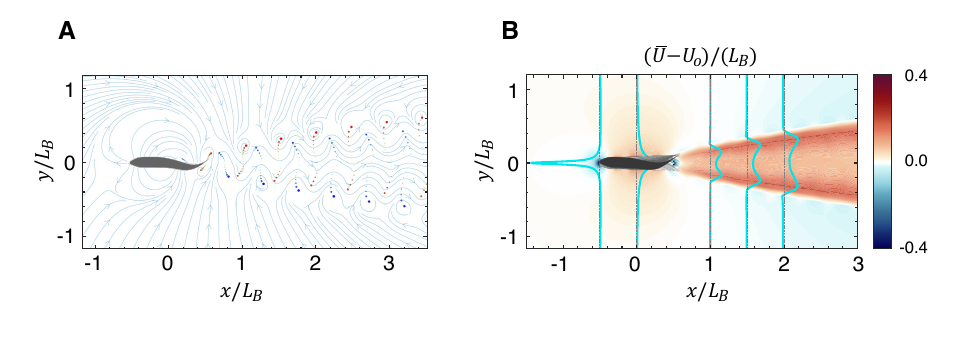}}
    \caption{A single swimming fish at steady-state with the potential field and the wake implemented. \textbf{A}: The streamline of the swimming fish. Dots behind the fish represent wake vortices. \textbf{B}: Time-averaged contour plot of $(\bar{U}-U_o)\/(L_B)$. Velocity profiles are plotted in cyan curves across the flow.}
    \label{fig:wakeIllustration}
\end{figure}

Fig. \ref{fig:wakeIllustration}. shows the streamline and velocity contour for the resultant velocity field of a solitary swimming fish. Both the potential body model and the wake model are implemented, as explained above. The qualitative similarity between this phenomenological flow model and the \textrm{DNS} data (Fig. \ref{fig:dns}) is worth noting. Similar to the DNS, the phenomenological model generates a symmetric dual-lobed ``jet" in the wake surrounded by a weak forward drafting velocity in the outer regions. Inline with the \textrm{DNS} predictions, the modeled wake generates an oblique pattern that slowly decays with distance. 

\begin{figure}[h]
    \centering
    {\includegraphics[width=1\textwidth]{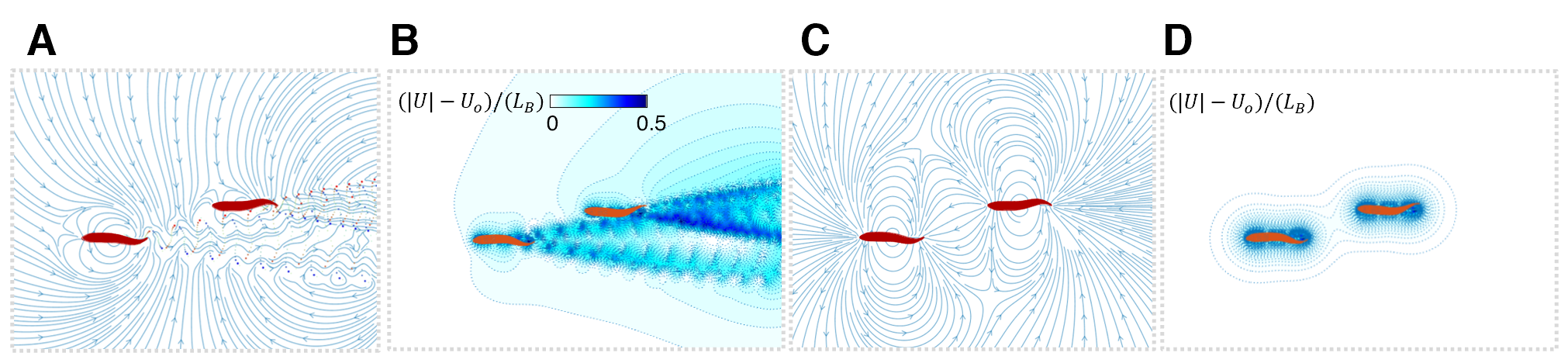}}
    \caption{Instantaneous streamline and velocity contour plots of a minimal school of fish (two fish). The relative position of the pair of fish is fixed. \textbf{A}: The streamline plot with the wake model and the potential field implemented. \textbf{B}: The contour plot of the induced velocity magnitude with the wake model. \textbf{C}: The streamline plot with the wake model disabled. \textbf{D}: The contour plot of the induced velocity magnitude with the wake model disabled.}
    \label{fig:streamlineWake}
\end{figure}

Fig. \ref{fig:streamlineWake} shows the flow field for a 2-fish school in the ``with wake" and ``without wake" models. The former corresponds to the vortex wake model being turned on in the simulation, and the latter is where the vortex wake model is turned off and only the potential flow due to the body is employed. We see clearly that while the potential flow disturbance due to the fish decays rapidly, the vortex wake extends far from the fish and the wake vortices from the leading fish interact strongly with the trailing fish. Thus, we expect that the vortex wake will significantly influence the schooling patterns that emerge from these models.

\subsection{Visual Sensing}
\begin{figure}[h]
    \centering
{\includegraphics[width=1\textwidth]{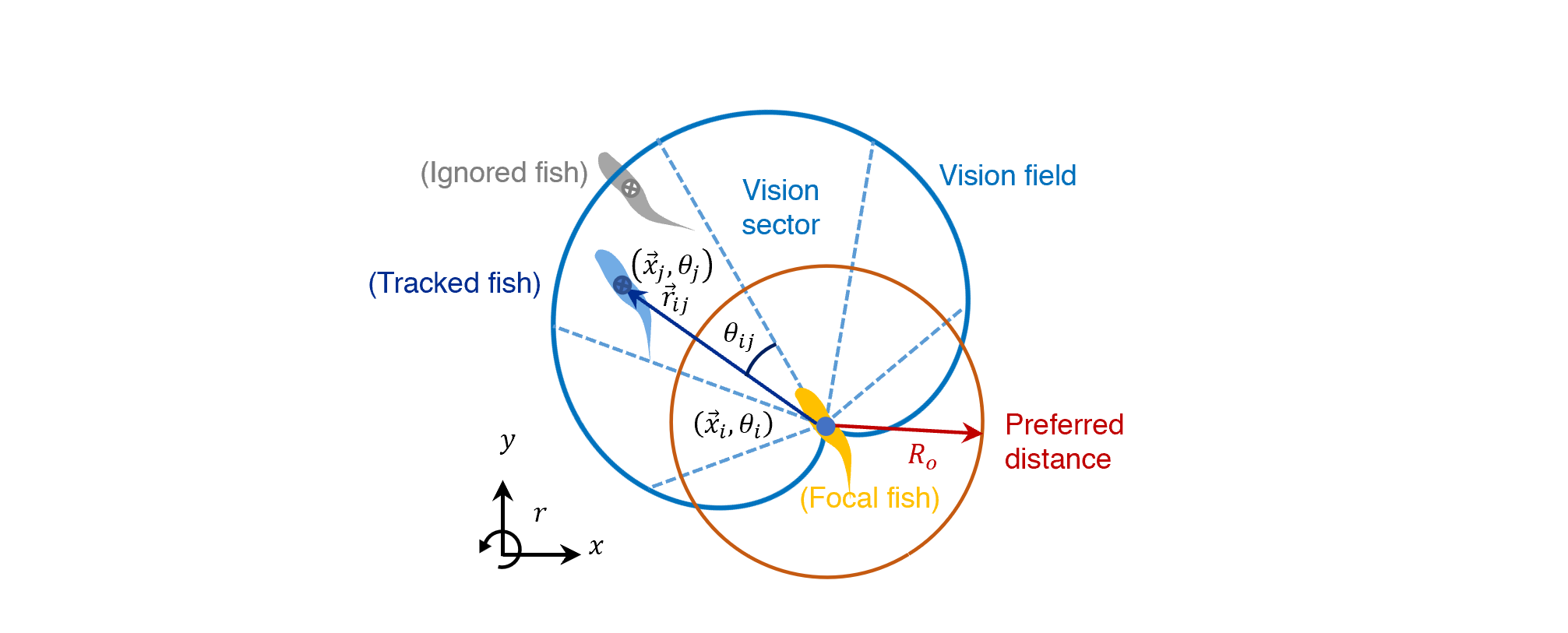}}
    \caption{Illustration of the preferred distance $R_o$ (red circle), and the cardioid vision field (region inside the blue curve) of fish $i$. The preferred distance and the visual field are scaled for better illustration.}
    \label{fig:visionField}
\end{figure}

Fish have a vision field that is biased in the forward (surge) direction \citep{newman2008effect}, and we employ a cardioid shaped vision field given by using ($\alpha \left(1 + \cos{\hat{\theta}_i}\right), \quad \text{where } \alpha = 2L_B$) \citep{calovi2014swarming,lopez2012behavioural,terzopoulos1994artificial} ( see Fig. \ref{fig:visionField}). We also incorporate ``attention parsimony \citep{terzopoulos1994artificial}" into the visual perception model, according to which a fish can track and respond to only a small number of targets within its visual field. Due to the lack of existing experimental references for mackerel fish, we conducted various tests and segmented the vision field into six sectors to accomplish this. We further assume that a focal fish detects only the closest fish within each sector since further away fish would be obstructed from its view. This rule of ``attention parsimony" is encoded in the model via the parameter $W_{ij}$ for each focal fish ``$i$" and tracked fish ``$j$." This parameter is set to unity if the tracked fish is the closest in each sector and set to zero otherwise.

\subsection{\label{sec:citeref}Social Interaction}
The terms $F_s$ and $T_s$  are the social interaction force and torque in the surge and yaw direction, respectively, and these are determined via considerations similar to those in many previous studies (\citep{zienkiewicz2018data,katz2011inferring,herbert2011inferring,escobedo2020data,parrish2002self}.  Each swimmer tries to maintain a preferred distance, $R_o$, from its neighbors (Fig. \ref{fig:visionField}), and when the distance between a tracked fish and the focal fish is larger/smaller than $R_o$, the focal fish will accelerate/decelerate in the surge direction and turn toward/away in the rotational direction to chase/avoid the tracked fish. These changes in speed and heading are accomplished by generating the surge force $F_s$ and yaw torque $T_s$ as per the model below.

The force due to social interaction in the surge direction, $F_s$, is assumed to be generated by the caudal fin. Given that generating such a force would take finite time, we employ a proportional-derivative (PD) controller to provide a smooth build-up and ramp-down of this force. The final expression for this force on the focal (``$i$") fish is as follows: 
\begin{equation}
    F_{s,i}=\frac{1}{\overline{\overline{W}}_i}\sum_{j=1}^{N_i}W_{ij}  \left( K_p(\vec{r}_{ij}\cdot \hat {\theta}_i-R_o)+K_d(U_{1,j}\cdot \hat{\theta}_i-U_{1,i}) \right) \left( \hat{r}_{ij}\cdot \hat{\theta}_i \right)\label{eq:8}
\end{equation}
where $N_i$ is the number of tracked fish within the active vision field of the focal fish at a given time-step, and $\overline{\overline{W}}_i=\sum_{j=1}^{N_i}W_{ij}$. Furthermore, $K_p$ is the attraction ``spring" constant in the surge direction, and $K_d$ is the coefficient for the velocity feedback term in the PD controller.

\begin{figure}[ht]
    \centering
    {\includegraphics[width=1\textwidth]{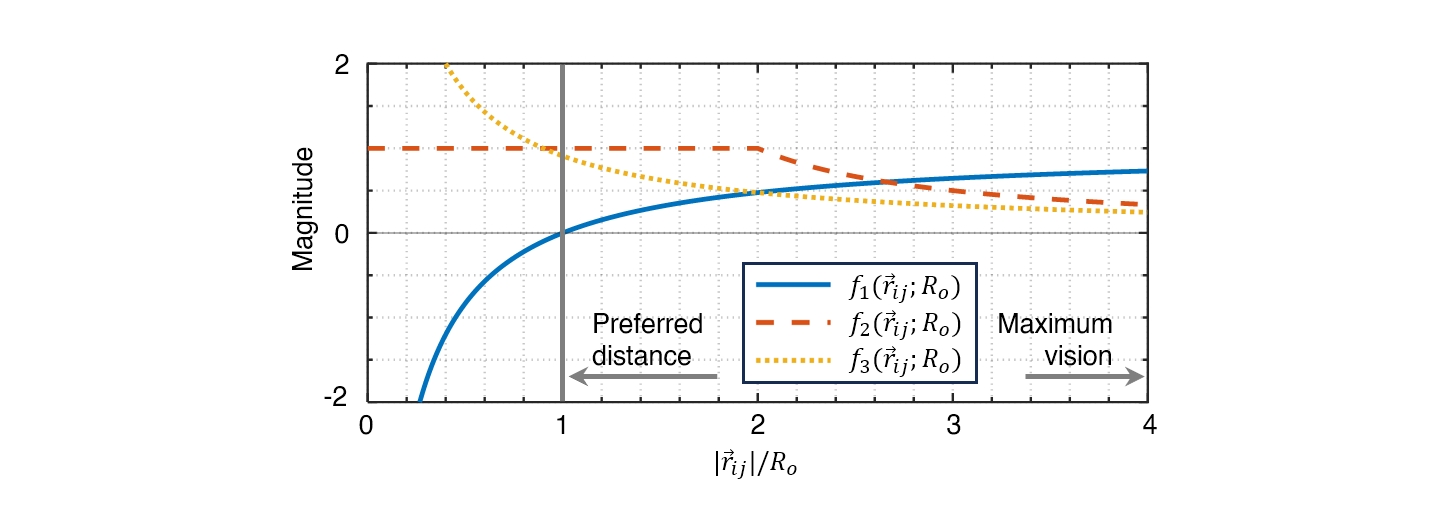}}
    \caption{The trends of the $f_1 (\vec{r}_{ij};R_o)$, $f_2 (\vec{r}_{ij};R_o)$, and $f_3 (\vec{r}_{ij^*};R_o)$, plotted against $|\vec{r}_{ij}|/R_o$. Positive values of the magnitude represent willingness to attraction and alignment, and negative values represent the opposite.}
    \label{fig:attalifunctions}
\end{figure}

In the yaw direction, fish are known to turn and align with tracked neighbors within their vision field, and this is considered key to the formation of schools (\citep{partridge1982structure, pavlov2000patterns, couzin2002collective}). We include two yaw torques connected with these social effects, $T_\textrm{att}$ and $T_\textrm{ali}$ (Eqn. \ref{eq:9}) where the former governs the propensity of fish to turn towards/away from a tracked fish and the latter controls the body orientation of the focal fish relative to that of the tracked fish. These two torques are formulated as follows:
\begin{equation}
    T_{\text{att}} = \frac{1}{\overline{\overline{W}}_i} \sum_{j=1}^{N_i} W_{ij} K_\textrm{AT}(\hat{\theta}_i \times \vec{\mathbf{r}}_{ij}) f_1(\vec{\mathbf{r}}_{ij};{R}_0); \quad
     T_{\textrm{ali}} = \frac{1}{\overline{\overline{W}}_i} \sum_{j=1}^{N_i} W_{ij} K_\textrm{AL} \left( \frac{\hat{\theta}_i}{2} \times \frac {\hat{\theta}_j}{2} \right) f_2(\vec{\mathbf{r}}_{ij}; {R}_0),
    \label{eq:9}
\end{equation}
where $f_1(\vec{\mathbf{r}}_{ij}; R_0) = \frac{|\vec{\mathbf{r}}_{ij}| - R_0}{|\vec{\mathbf{r}}_{ij}| + 0.1R_0}$ and $f_2(\vec{\mathbf{r}}_{ij}; R_0) = |\min \left[ \max \left( \frac{1}{|\vec{\mathbf{r}}_{ij}| - R_0}, -1 \right), 1 \right]|$ are attraction and alignment associated functions (see Fig. \ref{fig:attalifunctions}). 

Here, the attraction function $f_1$ is synthesized to induce a nearly uniform attraction outside the preferred separation and a repulsion force that increases rapidly inside the preferred separation for collision avoidance. The alignment function $f_2$ is designed to turn the fish away from a tracked fish that is closer than preferred separation and turn the fish towards a tracked fish more distant than the preferred separation.

The relative strength of attraction and alignment, which can be controlled by $\alpha_T=\frac{K_\textrm{AT}}{(K_\textrm{AT}+K_\textrm{AL})}$, is considered to be an important parameter for school formation and has been investigated in several previous studies \citep{calovi2014swarming,filella2018model,gautrais2012deciphering, zienkiewicz2018data}. In the current study, we have investigated the effect of this parameter by altering  $\alpha_T$ between high-alignment (low-attraction, $\alpha_T$=0.1) and high-attraction (low-alignment, $\alpha_T$=0.9).

The interaction between the focal fish and any boundary wall is also incorporated using a repulsion torque. The wall avoidance function $f_3$ prevents the fish from collision with the virtual tank by giving a relatively high turning near the wall. The effect of all the social forces and torques is limited to the extent of the vision field. Once a focal fish detects the wall in its vision range, it generates a repulsion torque, $T_\textrm{wall}=W_{ij} K_{\text{wall}} (\hat{\theta}_i \times \vec{\mathbf{r}}_{ij}^*) f_3(\vec{\mathbf{r}}_{ij}^*; R_o)$, where $f_3(\vec{\mathbf{r}}_{ij}^*; R_o) = \frac{R_o}{0.1R_o + |\vec{\mathbf{r}}_{ij}^*|}$, $j^*$ represents the index of the closest point on the wall (see Fig. \ref{fig:fishAnatomy}) from the focal fish, and $\vec{\mathbf{r}}_{ij}^*$ represents the normal vector towards outside of the boundary at the point $j^*$. 
\subsection{Final equations and Solution Procedure}
The final governing equations are obtained by incorporating the expressions of the hydrodynamic and social forces as well as torques into the right-hand-side of Eqns. \ref{eq:1}-\ref{eq:3}.  The definition and ranges (or values) of each parameter used in these expressions can be found in Table. \ref{table:parametersofwake}-\ref{table:parametersoffishwake}. All terms are nondimensionalized by the body length ($L_B$) of the fish, $L_B/U_o$ as the time scale, and $\rho_\textrm{water}$ as the density scale. The above equations are solved in MATLAB using the 4th-order Runge-Kutta time-advancement scheme with a time-step size of $\Delta t=0.05L_B/U_o$. 

In the current study, we employ two different domains for conducting simulations using our model: the first is a circular ``tank" with walls, which is employed to examine \emph{qualitative} schooling patterns that emerge from the model. Similar domains have been used in previous studies. The second domain is a large square domain whose area is $160\times N \times L^2_B$, where $N$ is the total number of fish. For this domain, periodic conditions are imposed in both directions for the Lagrangian (fish) and the Eulerian elements (induced flow velocity) in the model. This large domain with a size that grows proportionate to the size of the school and with periodic boundary conditions on all the boundaries, should allow for emergent dynamics that are relatively independent of confinement effects associated the external boundary. This second domain is used for a detailed \emph{quantitative} analysis of the effect of the vortex wake on the schooling patterns, which is the primary aim of the study.  In each simulation, the coordinates and orientation of each fish, as well as the coordinates and strengths of each of the vortices shed by the fish are tracked and used to determine the induced velocity and associated hydrodynamic forces and torques, as well as the social forces and torques on each fish.  We assume that the wake flow generated by a given fish is not modified by any trailing or adjacent fish. 

\subsection{Ensemble simulations of the collective motion}
\begin{figure}[ht]
    \centering
{\includegraphics[width=1\textwidth]{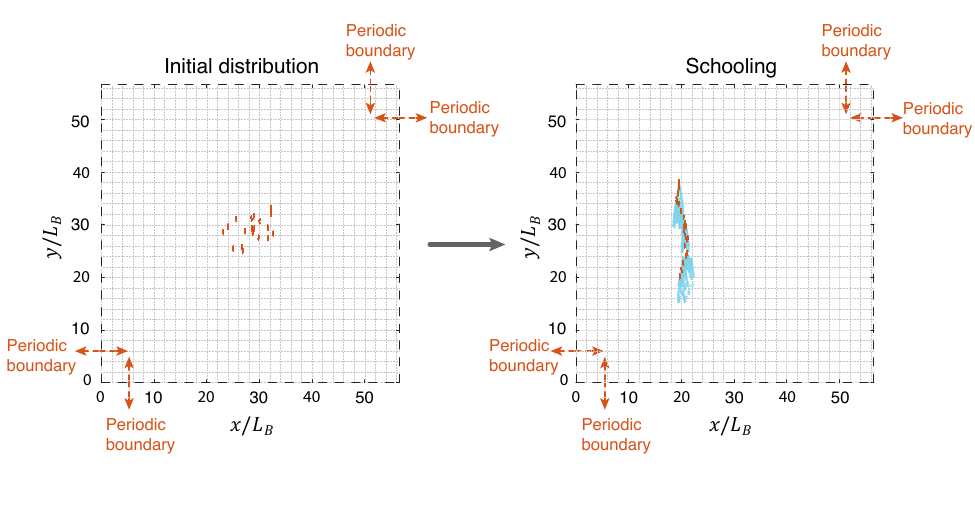}}
    \caption{Instantaneous distributions of a 20-fish, high-alignment ($\alpha_T$=0.1) schooling simulation, in the periodic domain. Plots illustrate fish distribution changes from the initial (left) to schooling (right) state. The blue dots represent generated wake vortices.}
    \label{fig:periodicDomain}
\end{figure}

The equations governing the motion of the fish are non-linear and the interaction between multiple fish results in solutions that exhibit a sensitivity to initial conditions that is typical of all such non-linear dynamical systems. To obtain reliable statistics of the behavior and topologies that emerge from these interactions, we conduct an ensemble of 300 simulations for every case in the following part. In each simulation, the fish locations are initialized randomly in the central region of the square domain (Fig. \ref{fig:periodicDomain}). For different numbers of fish, the density of fish is maintained at 1 fish per 160 $L^2_B$ by adjusting the edge length of the domain accordingly. Each simulation is allowed to reach a stationary state as determined by the swimming velocity of the group, and this can take up to 20,000 time-steps. Analysis of the emergent collective behavior is based on statistics and principal component analysis (\textrm{PCA}) in this stationary state and the details of the \textrm{PCA} process are provided in Appendix \ref{sec:appendixA}.

\section{Results}
\subsection{ Emergent patterns of collective motion}
\begin{figure}[h]
    \centering
{\includegraphics[width=1\textwidth]{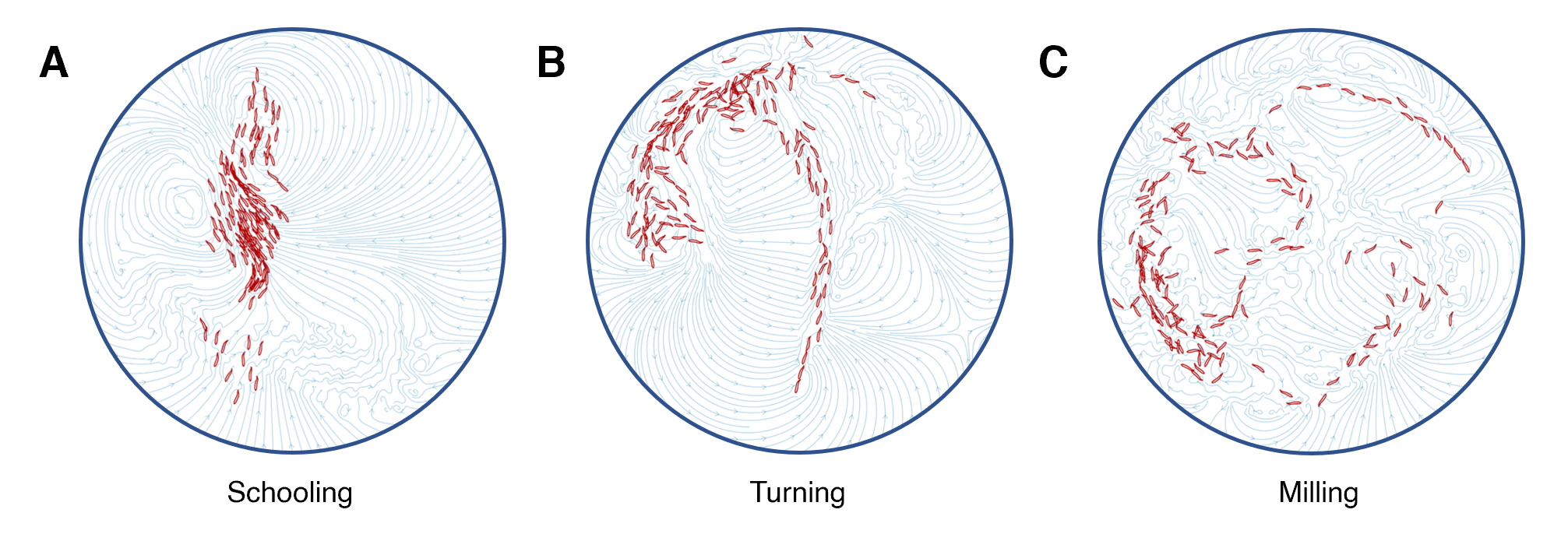}}
    \caption{Instantaneous plots of three distinguish collective patterns of 150 fish swimming in a circular tank, streamlines annotated. \textbf{A}: Schooling, $\alpha_T=0.1.$ \textbf{B}: Turning, $\alpha_T=0.5.$  \textbf{C}: Milling, $\alpha_T=0.9.$}
    \label{fig:fishPatterns}
\end{figure}

In Fig. \ref{fig:fishPatterns}, we show the emergence of complex schooling topologies and kinematics from our agent-based model, and we provide some qualitative data on the types of patterns. The relative strength between attraction and alignment (denoted by $\alpha_T$ in the current model) is known to be an important parameter in the emergent behavior of these modeled collective motions \citep{calovi2014swarming,filella2018model,gautrais2012deciphering, zienkiewicz2018data}. In high-alignment situations ($\alpha_T=0.1$), fish swim in the same direction as their neighbors, resulting in a more polarized pattern. In contrast, in the high-attraction situation ($\alpha_T=0.9$), fish prioritize swimming towards their neighbors, even though their swimming directions are not aligned, and they tend to form localized milling patterns. The schooling, turning, and milling (Fig. \ref{fig:fishPatterns}) can emerge when the number of fish is sufficiently large. Those resultant patterns qualitatively match with some previous computational models (\citep{couzin2002collective,newman2008effect,tunstrom2013collective,calovi2014swarming,filella2018model}). 
\subsection{Quantitative Analysis of the Impact of Wakes}
In this section, we examine the impact of the wake on the collective motion of fish. We have conducted multiple sets of simulations, varying two essential features of fish schools: the number of fish (5, 10, 15, and 20) and the social effect weights so as to generate two behavior conditions, high-alignment: ($\alpha_T= 0.1$) and high-attraction: ($\alpha_T= 0.9$). We also simulate a control group with identical settings but without the wake.

\begin{figure}[h]
    \centering
    {\includegraphics[width=1\textwidth]{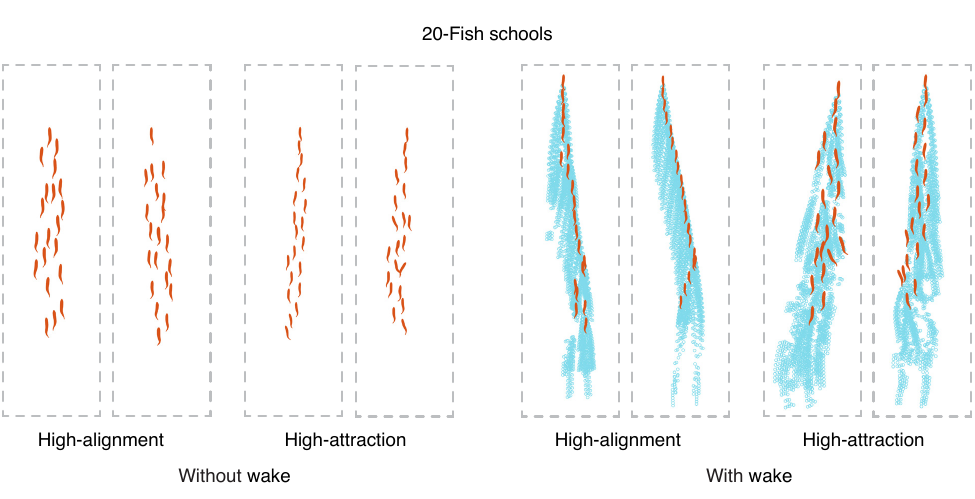}}
    \caption{Instantaneous distributions of 20-fish schools, comparing in different wake and behavior modes. Blue dots represent fish wake vortices. Each dashed rectangular box denotes an individual representative.}
    \label{fig:caseIllutration}
\end{figure}

Fig. \ref{fig:caseIllutration} shows representative snapshots of 20 fish under high-alignment and high-attraction conditions. In the high-alignment mode, the 20-fish schools indicate that adding the wake arranges the school mostly along a localized ``diamond" pattern, with each follower located along the edge of the wake of the preceding fish. This phenomenon can be explained by the narrow drifting region near the edge of the wake (see Fig. \ref{fig:wakeIllustration}), which induces a forward force to attract trailing fish to stay near the edge passively. 

To understand the formation of this unique layout, we start with the first pair of fish at the head of the school. If the trailing fish stays near the right edge of the wake generated by the leading fish, the wake region behind the trailing fish would consist of one superposition of their wake's right edges and two left edges, which are separate from each other. Hence, the enhanced right edge would have a stronger induced velocity to attract the following trailing fish to stay around it, forming a stronger right edge. Furthermore, the high-alignment mode is another key feature that helps form such patterns. The high-alignment behavior mode ensures individuals' bias in putting themselves in the same swimming direction as their targeted neighbors. Therefore, Each fish's localized wake can be successfully superposed together and consistently transmitted downstream. The high-alignment mode combined with the favorable drifting regions of the wake results in the distinguishingly local ``diamond" pattern. This finding supports the hypothesis that there are hydrodynamic favorable regions outside the wake \citep{weihs1973hydromechanics}. 

However, in the high-attraction condition, the shape of schools of ``with wake" is not qualitatively distinct from ``without wake" ones because the high-attraction nature endows individuals to dynamically approach targeted neighbors, which fails in forming consistent relative positions among streamwise fish. Hence, wakes can hardly be superposed, and their passively induced hydrodynamic benefits have to compromise on the behavior's desires. 
\begin{figure}[ht]
    \centering
{\includegraphics[width=1\textwidth]{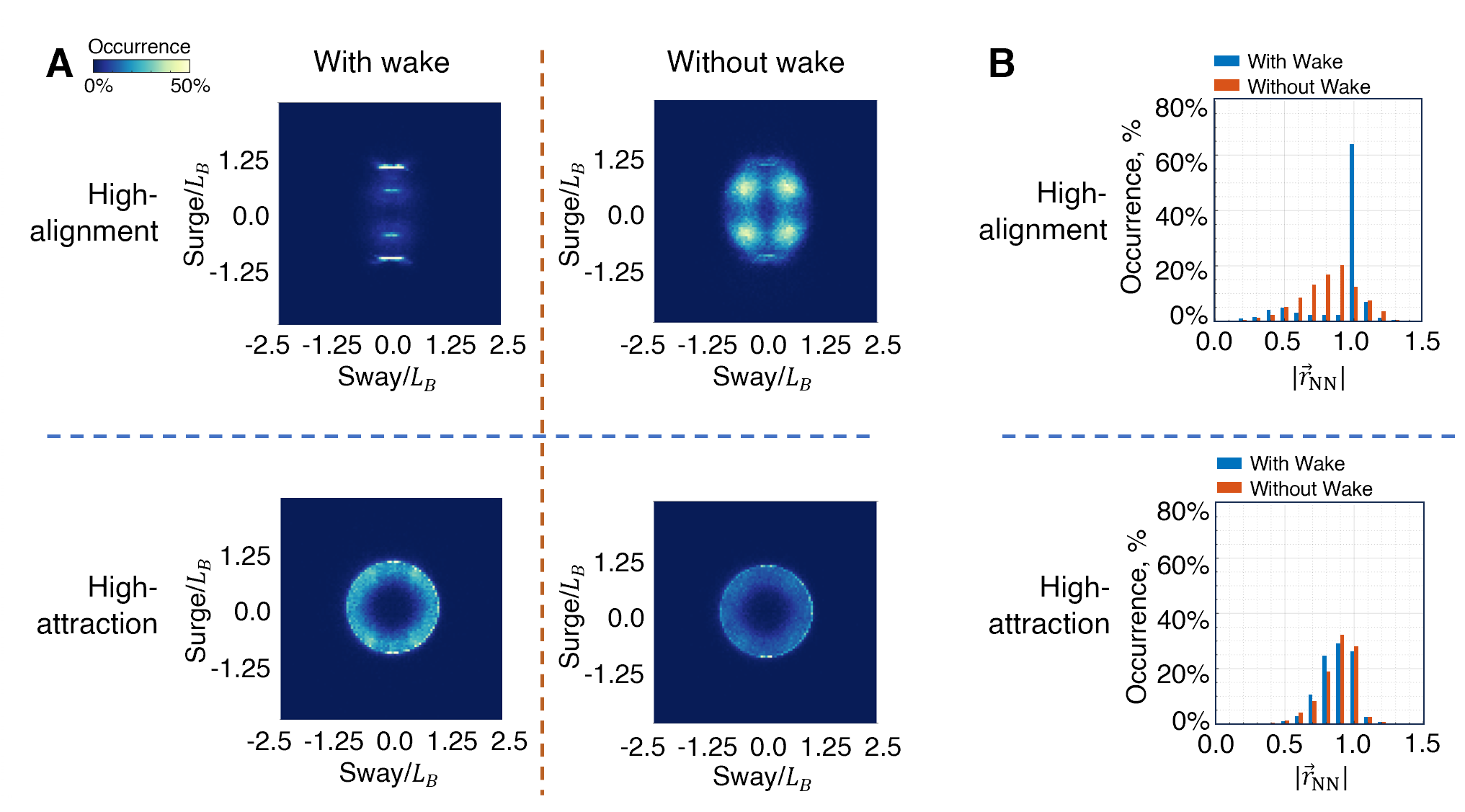}}
    \caption{Statistics of the nearest neighbor distance $\text{(NND)}$ for 20-fish cases, comparing between ``with wake" and ``without wake" conditions. \textbf{A}: The high-alignment and high-attraction conditions are examined using 300 ensemble simulations. The occurrence is normalized by the highest value on each heat contour plot. \textbf{B}: The $|\vec{r}_\text{NN}|$ is the distance between the nearest neighbor to the focal fish.}
    \label{fig:20Fish-Ro}
\end{figure}

Another metric for quantifying the organizational structure of fish schools is the nearest-neighbor-distance (\textrm{NND}), which has been widely used in the literature. The \textrm{NND} usually quantifies the distance between each focal fish and the nearest neighbor, but we also include the relative angle. Fig. \ref{fig:20Fish-Ro}\textbf{A} shows the distribution of the \textrm{NND} for 20-fish schools for the same values of $\alpha_T$ as the previous section, high-alignment: ($\alpha_T= 0.1$) and high-attraction: ($\alpha_T= 0.9$). Considering the ``without wake" case, we see that in the high-attraction situation, the fish are clustered very close to the preferred distance $R_o$ but do not have a preferred angular placement. A more well-defined structure emerges for the high-alignment case, with fish clustered in four symmetric regions relative to each other. 

Comparison with the ``with wake" case shown in Fig. \ref{fig:20Fish-Ro}\textbf{A}  indicates that the presence of the wake has a significant effect on the \textrm{NND} and the relative placement of the fish around each other. The difference is readily apparent for the high-alignment case, where the presence of the wake seems to force the fish into more of a streamwise aligned configuration. The span of the distribution matches our analysis above that the wake from the leading fish passively attracts the focal fish near the edge in the high-alignment case.

Fig. \ref{fig:20Fish-Ro}\textbf{(B)} shows the \textrm{PDF} of the \textrm{NND} for the four scenarios shown in \ref{fig:20Fish-Ro}\textbf{(A)}, and this confirms that the effect of the wake increases as fish shift from high-attraction to high-alignment. For the high-attraction case, we find that while in ``with wake," over 60\% of the fish are located at the preferred distance of $R_o$, in ``without wake," there is a wider range of preferred distances. This confirms that the wake tends to introduce more order in the fish school topology, especially when the social alignment force is dominant. \textrm{NND} provides some useful information about the effect of social forces and wake effects on the cohesion of the school. Still, \textrm{NND} assumes that the schools are homogeneous, and it is therefore unable to provide any information about the overall shape and topology of the school. For this reason, we turn to the principal component analysis (PCA) based analysis. The detailed conduction of PCA is collected in Appendix \ref{sec:appendixA}.
\begin{figure}[ht]
    \centering
{\includegraphics[width=1\textwidth]{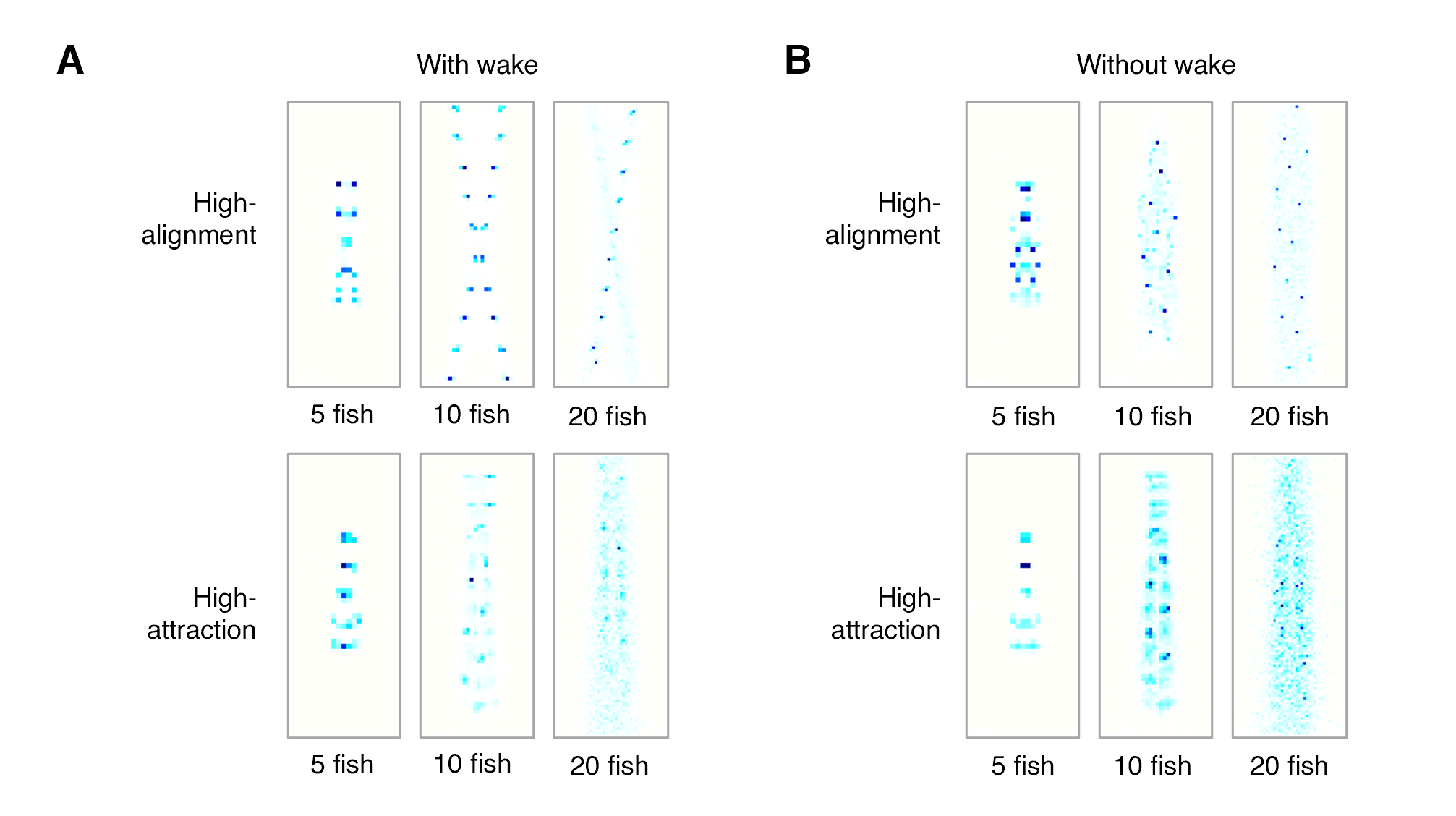}}
    \caption{Reconstructed school distribution from first principal components of different numbers of fish at high-alignment and high-attraction, respectively. Plots are scaled to provide a better view of the area of the collective group of fish. The actual sizes of each set of the number of fish are $(\pm1.41, \pm3.54) L_B$, $(\pm2.00, \pm5.00) L_B$, and  $(\pm2.83, \pm7.07) L_B$ for 5 fish, 10 fish, and 20 fish, respectively. The origin represents the centroid of the school. \textbf{A}: Both the potential field and the wake are implemented. \textbf{B}: The potential field is implemented while the wake is disabled.}
    \label{fig:PCAWake}
\end{figure}

Figure \ref{fig:PCAWake} shows the reconstructed school distribution from the first principal components (PCs) for various schooling conditions under ``with wake" and ``without wake" cases. Fig.~\ref{fig:PCAWake}\textbf{A} illustrates the spatial arrangement of schools in high-alignment and high-attraction cases for schools consisting of 5, 10, and 20 fish. Under the ``with wake" condition, the schools exhibit a highly structured topology, particularly in the high-alignment cases, where the fish form well-defined oblique (diagonal) patterns along the axial direction. This arrangement is indicative of hydrodynamic interactions influencing school formation. Conversely, under the ``without wake" condition, the schools show a more dispersed and less organized topology, with no clear oblique structures emerging even at a small school size.

The high-attraction cases in Fig.~\ref{fig:PCAWake}\textbf{A} demonstrate similar trends, but the oblique patterns are less pronounced compared to the high-alignment cases. This difference can be attributed to the relative strength of attraction forces versus alignment forces. In the high-attraction cases, the fish are more strongly influenced by inter-individual forces, which dominate over the hydrodynamic effects of the wake, leading to a denser but less structured school topology. These results emphasize the critical role of wake-induced hydrodynamic forces in shaping the school's spatial organization, particularly when alignment dominates over attraction.

Fig.~\ref{fig:PCAWake}\textbf{B} supports these observations by presenting the same cases but in the absence of wake effects. The schools are generally more symmetric about the axial centerline and exhibit reduced spatial diversity, with no clear oblique patterns regardless of the schooling condition or fish number. This comparison highlights the influence of wake dynamics on schooling behavior, especially in larger groups where hydrodynamic interactions become increasingly significant.

\begin{figure}[ht]
    \centering
    {\includegraphics[width=1\textwidth]{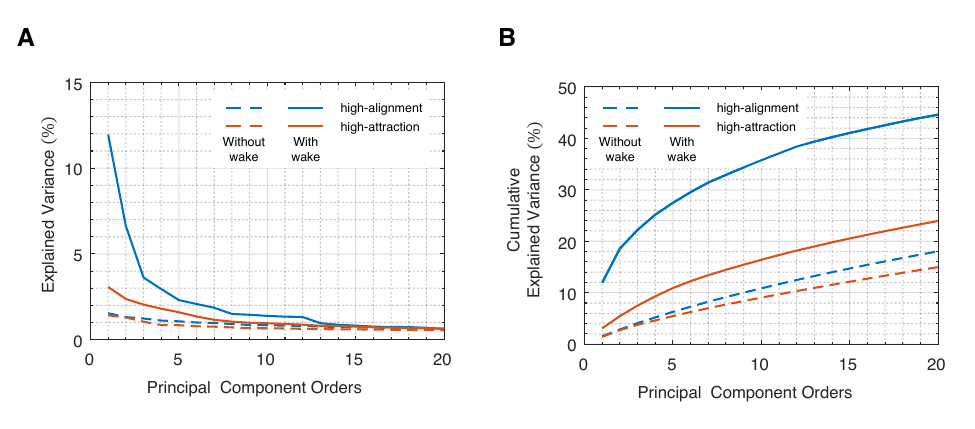}}
    \caption{PCA results on 10-fish cases at high-alignment and high-attraction conditions. \textbf{A}: Explained variance, and \textbf{B}: corresponding accumulative percentages of the first 20 principal components.}
    \label{fig:explainVarianceWake}
\end{figure}

Moving to the quantitative analysis, Fig.~\ref{fig:explainVarianceWake} provides further insights into the impact of wake dynamics on schooling topology by examining the explained variance of the PCs. In Fig.~\ref{fig:explainVarianceWake}\textbf{A}, the first PC accounts for a significantly larger portion of the variance in the ``with wake" condition compared to the ``without wake" condition. For high-alignment cases, the first PC explains 11.95\% of the variance "with wake," compared to only 1.54\% "without wake." Similarly, for high-attraction cases, the first PC accounts for 3.07\% of the variance "with wake," compared to 1.42\% "without wake." These differences indicate that wake dynamics introduce a stronger organizing effect in the schools, particularly in high-alignment cases.

The cumulative explained variance in Fig.~ \ref{fig:explainVarianceWake}\textbf{B} further reinforces these findings. The first 20 PCs in the ``with wake" condition explain 44.60\% and 23.94\% of the total variance for high-alignment and high-attraction cases, respectively, compared to only 18.05\% and 14.97\% in the ``without wake" condition. This demonstrates that the wake's influence introduces a more distinct and structured topology, resulting in a higher degree of spatial coherence within the school.

Interestingly, the presence of the wake induces a regularizing effect on the school topology, counterintuitive to the expectation that wake vortices might increase disorder. This effect is particularly evident in high-alignment cases, where the hydrodynamic instability caused by the wake pushes trailing fish away from the wake centerline of the leading fish. At the same time, alignment forces prevent the fish from moving too far apart, resulting in the observed oblique patterns. For high-attraction cases, the increased influence of inter-individual forces reduces the dominance of wake effects, leading to less pronounced oblique patterns but still enhancing the overall spatial coherence compared to ``without wake" conditions.

In summary, Figs.~\ref{fig:PCAWake} and \ref{fig:explainVarianceWake} demonstrate that wake-induced hydrodynamic interactions significantly shape schooling topology, particularly under high-alignment conditions. The results highlight the interplay between hydrodynamic forces, alignment, and attraction, offering insights into schooling behavior. Next, we examine specific wake characteristics, including spanwise width and circulation strength, to better understand their role in shaping schooling patterns and dynamics.

\subsection{Impact of wake characteristics on schooling patterns}
\begin{figure}[ht]
    \centering
{\includegraphics[width=1\textwidth]{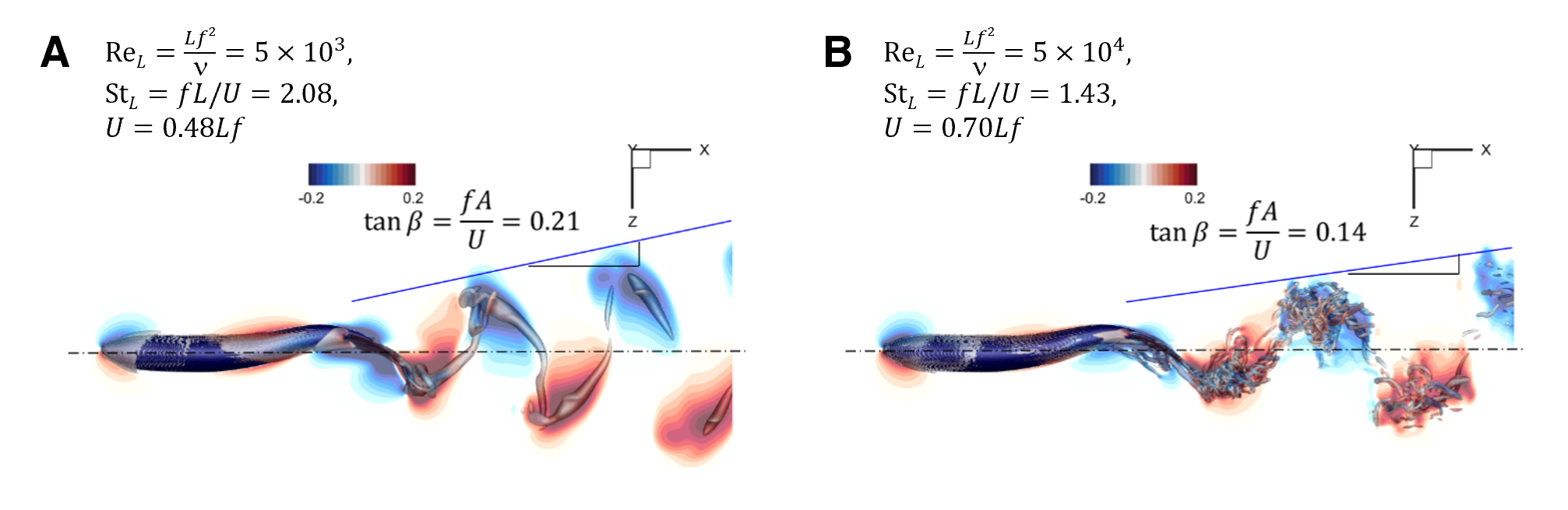}}
    \caption{Direct numerical simulations of a free-swimming fish under different $\mathbf{Re}_L(\mathbf{St}_L)$. \textbf{A}: $\mathbf{Re}_L=5\times 10^3;\, \mathbf{St}_L=2.08$. Results from our previous study \citep{seo2022improved}. \textbf{B}: $\mathbf{Re}_L=5\times 10^4;\, \mathbf{St}_L=1.43$. The same approach and settings as \textbf{A}.}
    \label{fig:StRe}
\end{figure}

\begin{figure}[hbt!]
    \centering
{\includegraphics[width=1\textwidth]{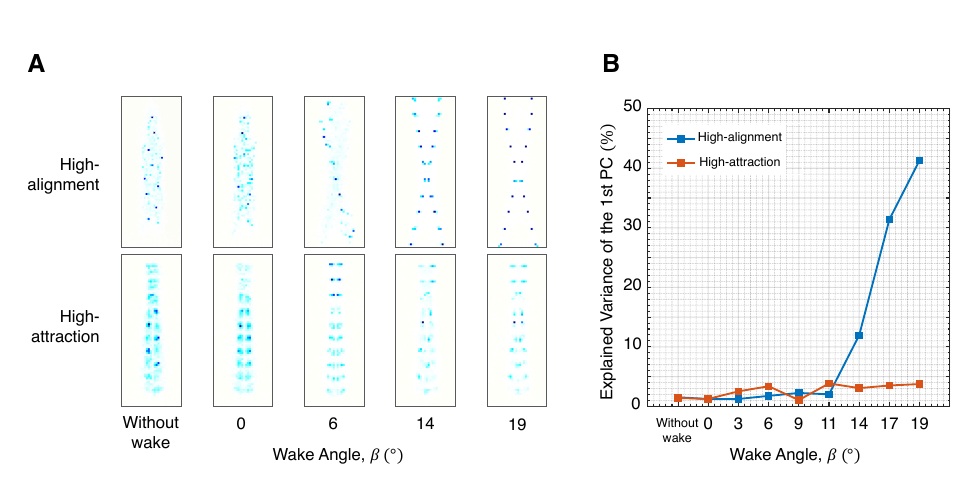}}
    \caption{Results of PCA on fish distributions of 10-fish schools with various wake width (angle), $N=1,500$. Simulations are conducted at high-alignment and high-attraction, respectively. \textbf{A}: Reconstructed distribution from the first principal component of the PCA. The size of each representative is $(\pm2.00, \pm5.00) L_B$. \textbf{B}: The portion of the first principal components of PCA on fish distributions.}
    \label{fig:PCAVs}
\end{figure}

The width of the wake in the spanwise direction can be altered by the aspect ratio ({\AR}) of the caudal fin and the Reynolds number ($\mathbf{Re}$). A previous study \citep{dong2006wake} indicated that different {\AR} values determine the spanwise width of the trailing wake. Smaller {\AR}  values result in wider wakes, whereas larger {\AR}  values lead to narrower and longer wakes. Our \textrm{DNS} of a single fish with different $\mathbf{Re}$ indicates that higher $\mathbf{Re}$ can result in narrower wakes as well (Fig. \ref{fig:StRe}). Therefore, by simulating fish schooling with varying wake widths, we can assess the influence of different {\AR}  and $\mathbf{Re}$ on the terminal schooling pattern. In this model, we manipulated the wake width by adjusting the drifting speed ($V_s$) of wake vortices. Changing $V_s$ results in corresponding changes in wake angle, $\tan \beta=V_s/U_1$. The default wake angle is $14^\circ$. Using the same PCA pipeline, we conducted ensemble simulations with groups of 10, exploring both narrower and wider wake lengths while keeping the strength of the vortices constant. The first PCs for each case are presented in Fig. \ref{fig:PCAVs}\textbf{A} and the corresponding explained variance portions are shown in Fig. \ref{fig:PCAVs}\textbf{B}.

In the 10-fish scenario, apparently in Fig. \ref{fig:PCAVs}, wider wakes greatly contribute to organizing individuals within the school under the high-alignment mode while having little effect in the high-attraction mode. On the opposite, when wakes get narrower, there are no distinct differences in fish topology between the high-alignment and high-attraction sets. This observation can also be found statistically, as Fig. \ref{fig:PCAVs}\textbf{B} shows the first \textrm{PC} of each set of various wake angles. When $\beta \leq 11^{\circ}$, wider wakes have a detrimental effect on maintaining a clear schooling pattern in the high-alignment mode. However, starting from $\beta = 14^{\circ}$, wider wake widths provide a more pronounced advantage in the high-alignment mode. Consequently, fish that generate wider wakes due to a low $\mathbf{Re}$ or small {\AR}  of the caudal fin benefit from hydrodynamics in schooling organization when they are highly polarized, corresponding to the high-alignment mode in our results. In contrast, hydrodynamics slightly influence schooling patterns for high-attraction schools, regardless of the caudal fin {\AR} or the effects from $\mathbf{Re}$ on the wake width.

\subsection{Impact on schooling pattern from hydrodynamics: Strength of the vortices}
\begin{figure}[H]
    \centering
{\includegraphics[width=1\textwidth]{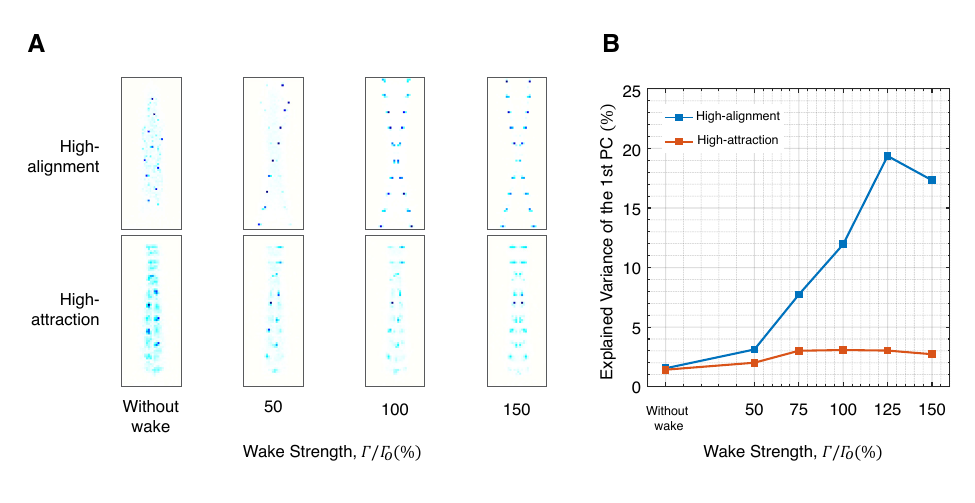}}
    \caption{Results of PCA on fish distributions of 10-fish schools with various wake strength, $N=1,500$. Simulations are conducted at high-alignment and high-attraction, respectively. \textbf{A}: Reconstructed distribution from the first principal component of the PCA. The size of each representative is $(\pm2.00, \pm5.00) L_B$. \textbf{B}: The portion of the first principal components of PCA on fish distributions.}
    \label{fig:PCAStrength}
\end{figure}
The strength of the vortices directly correlates with the extent of hydrodynamic influence experienced by fish. To examine the benefits of hydrodynamics for fish groups, we conducted PCA using the same pipeline as previous sections on schooling simulations under various vortex strengths, ranging from 50\% to 150\% of the default value. Fig. \ref{fig:PCAStrength} illustrates the dominant patterns (first PC) of each case. It is evident that higher vortex strength results in clearer patterns when fish are in the high-alignment mode but does not necessarily lead to clearer patterns in the high-attraction mode. As shown in Fig. \ref{fig:PCAStrength}\textbf{B} for 10-fish schools, increased vortex strength results in more organized patterns, particularly in the high-alignment mode. However, the peak in the high-alignment mode indicates that excessively strong hydrodynamic effects could lead to less organized patterns. Through simulations involving fish generating different strengths of wake vortices, we observed that hydrodynamics from the wake is beneficial for organizing the distribution of individuals when they are highly polarized. An optimal level of vortex strength is expected to yield the best results.

Consequently, our evaluation of the impact on fish schooling patterns from hydrodynamics considered varying wake widths (corresponding to different caudal fin {\AR} or $\mathbf{Re}$) and wake strengths. We observed significant contributions from the wake, emphasizing its importance in fish schooling. Interestingly, in high-attraction schools, the influence of hydrodynamics is reduced due to less organized wake patterns and the rapid locomotion of swimmers' relative positions.

\section{\label{sec:conclusion}Conclusion}

This study presents a novel agent-based model of fish schooling that integrates social behavior, vision-based sensing, and, crucially, the hydrodynamic forces induced by vortex wakes. The inclusion of wake-induced hydrodynamic interactions sets this model apart from classical schooling models, enabling the exploration of how these forces influence school topology and dynamics. The results demonstrate that wake-induced interactions can promote the emergence of more ordered schooling patterns, such as the diagonal formations observed in high-alignment cases. While such patterns are not commonly observed in nature, their emergence reflects an optimal solution driven by hydrodynamic efficiency within the model. This highlights the balance between hydrodynamic optimization and other behavioral factors, such as social cohesion and predator evasion, which likely dominate natural schooling behavior.

Wake characteristics and vortex strength significantly influence schooling patterns. Wider wakes, associated with lower Reynolds numbers (\( \mathbf{Re} \)) or smaller caudal fin aspect ratios ({\AR}), enhance schooling organization in highly polarized groups (high-alignment) while smaller impact on less polarized schools (high-attraction). Similarly, stronger vortices promote more statistically dominant schooling patterns in high-alignment modes, but excessive strength can destabilize the topology. These results emphasize the nuanced role of hydrodynamic interactions in shaping fish schools, particularly under conditions of high polarization.

By parameterizing species-specific features, including size, shape, and social interaction preferences, the model demonstrates flexibility in simulating a range of schooling behaviors under various environmental conditions. Unlike prior models that focus on reproducing experimentally validated schooling patterns, this study primarily aims to reveal the hydrodynamic effects on fish topology. The insights gained here complement experimental studies by offering a hydrodynamics-first perspective on schooling behavior, which can guide future explorations into the interplay between social and physical factors.

Despite its restriction to two dimensions, the model captures many essential features of schooling dynamics, particularly in the context of hydrodynamic interactions. The demonstrated sufficiency of 2D modeling for many schooling behaviors supports its use as a first-order approximation. Extensions to three dimensions in future work could provide a more detailed understanding of complex schooling structures. Additionally, while the current model incorporates only vision-based sensing, additional sensory modalities, such as pressure and flow sensing, could enhance its ability to simulate natural fish interactions.

In conclusion, this study provides a versatile framework for investigating the hydrodynamic contributions to collective fish behavior. By focusing on the hydrodynamic effects on schooling topology, the model is a valuable tool for researchers exploring the physical mechanisms underlying schooling behavior. Its flexibility and ability to simulate species-specific behaviors make it well-suited for future studies on collective behavior, bio-inspired applications, and the role of hydrodynamic forces in shaping group dynamics.

\begin{acknowledgments}
This work is supported by ONR Grants N00014-22-1-2655 and N00014-22-1-2770 monitored by Dr. Robert Brizzolara. We acknowledge useful discussions with Drs. Matt McHenry (University of California at Irvine) and Eva Kanso (University of Southern California).
\end{acknowledgments}

\section*{Data Availability Statement}
The data that support the findings of this study are available from the corresponding author upon reasonable request.

\appendix
\section{\label{sec:appendixA}Appendix}

\subsection{Principal Component Analysis}

Metrics like polarity \citep{parrish2002self,filella2018model}, which quantifies the degree of co-orientation, and mean nearest neighbor distance (\textrm{NND}), which measures school compactness, are commonly used to analyze schooling behavior. While these scalar metrics provide valuable insights, they cannot capture the topology or spatial structure of the school. For example, if orientation and spacing are identical, they cannot distinguish between a phalanx (fish swimming abreast) and an inline configuration (fish aligned one behind the other). 

\begin{figure}[H]
    \centering
    {\includegraphics[width=1\textwidth]{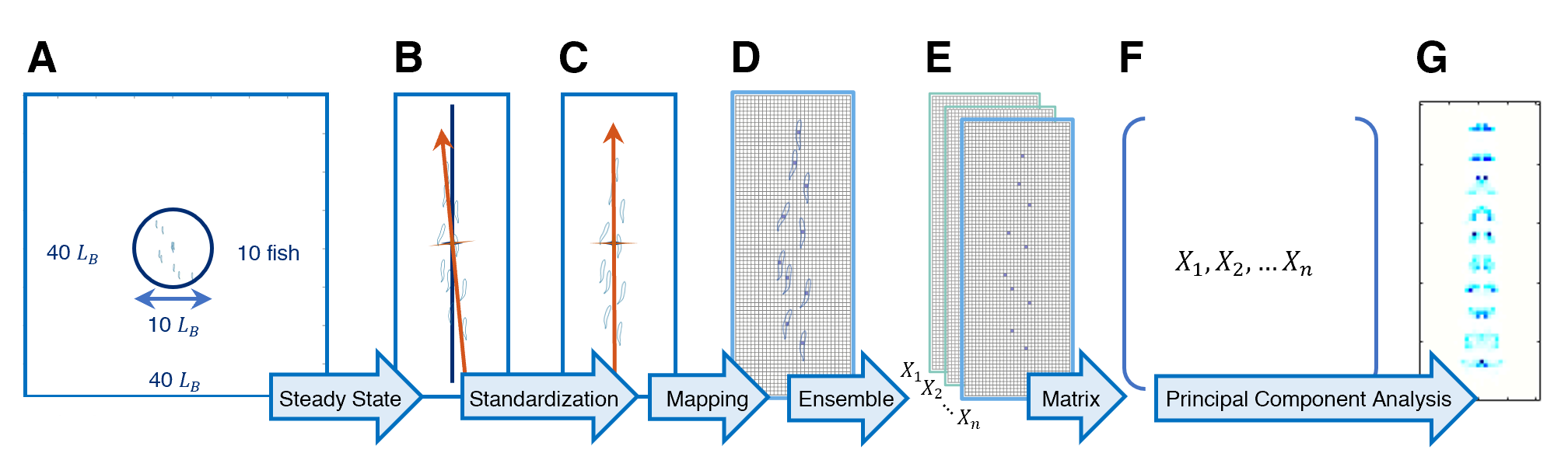}}
    \caption{Workflow for conducting Principal Component Analysis (\textrm{PCA}) on 10-fish schooling distributions. The same procedure is applied to schools with different fish numbers, with dimensions scaled to maintain consistent fish density. \textbf{A}: Initial random distribution of fish within a circular area (diameter: 10 $L_B$) at the center of a square domain (side length: 40 $L_B$) with periodic boundaries. \textbf{B}: Fish school reaches a steady state, and the school centroid and mean heading direction (red arrow) are calculated. \textbf{C}: Fish positions are standardized based on the centroid and mean heading direction. \textbf{D}: The standardized positions are mapped onto a binary pixel grid, where pixels containing fish are assigned a value of 1, and others are 0. \textbf{E}: Ensemble simulations (300 instances) are conducted with 5 steady-state snapshots extracted per simulation. \textbf{F}: A matrix is constructed from 1,500 pixel grids (300 simulations $\times$ 5 snapshots). \textbf{G}: \textrm{PCA} is applied to the matrix to identify dominant schooling patterns. The first principal component represents the most common fish distribution and is mapped back onto the pixel grid for visualization.}

    \label{PCAProcess}
\end{figure}

To analyze topological features, we employ Principal Component Analysis (\textrm{PCA}), a method widely used in pattern recognition and dimensionality reduction \citep{jollife2016principal}. \textrm{PCA} allows for both qualitative and quantitative characterization of school organization by identifying dominant patterns in fish distributions. The procedures used in this study are outlined in Fig.~\ref{PCAProcess}.

In this study, each simulation was run until the school reached a stationary state. At this point, five snapshots were extracted at sufficiently spaced time intervals. Fish positions and orientations were standardized by subtracting the centroid and aligning to the mean heading direction (Fig.~\ref{PCAProcess}\textbf{B}, \textbf{C}). The standardized positions were mapped onto a two-dimensional pixel board (Fig.~\ref{PCAProcess}\textbf{D}), and the pixel boards from all simulations and snapshots were compiled into a ($N_\textrm{pixels} \times 1,500$) matrix (Fig.~\ref{PCAProcess}\textbf{F}). PCA was applied to this matrix to identify principal components and their associated eigenvalues. The first principal component (PC), representing the most dominant schooling pattern, was visualized on the pixel board to provide insights into the school topology.

\subsection{Model Parameters}
Tables of the model parameters used in this study are included below for reference. These tables detail fish physiology, social behavior parameters, and hydrodynamic factors. The rich parameter space allows this model to fit different fish species with corresponding behavior and hydrodynamic features.
\renewcommand{\arraystretch}{1}
\begin{table}[H]
\caption{\label{tab:table4}Parameters of fish physiology and social interactions.}
\scriptsize 
\begin{adjustbox}{width=0.7\textwidth,center}
\begin{tabular}{ll}
\toprule 
\textbf{Parameter} & \textbf{Expression}\\ \hline
Fish body length & \(L_B = 0.1 \, \text{m}\) \\
Dimensions of prescribed elliptic body shape & \(a = 0.5 L_B; \, b = 0.075 L_B\) \\
Cross-sectional area & \(A_{\text{cross}} = (2b)(2c)\) \\
Side area & \(A_s = (2a)(2c)\) \\
Fish volume & \(\forall_{\text{fish}} = (2a)(2b)(2c)\) \\
Fish density & \(\rho_{\text{fish}} = 1000 \, \textrm{kg/}m^3\) \\
Fish mass & \(M = \rho_{\text{fish}} \forall_{\text{fish}}\) \\
Fish moment of Inertia & \(I = \frac{M_{\text{fish}}}{5} (a^2 + b^2)\) \\
Fish equilibrium speed & \(U_o = 1L_B/s\) \\
Fish base-tailbeat frequency \citep{bainbridge1958speed} & \(\mathcal{F} = \frac{4}{3}(U_o/L_B + 1)\) \\
Damping coefficient & \(c = \mathcal{F}\) \\
Limit of \(F_s\) & \(W_{\text{propulsion}} = 10\) \\
Vision field & \(\alpha = 2L_B; \alpha(1 + \cos(\hat{\theta}_i))\) \\
Visual sectors & \(N_{\text{sector}}; \theta_{\text{sector}} = {2\pi}/{N_{\text{sector}}}\) \\
Number of neighbors in vision and tracked & \(n_{\text{tracked}}\) \\
Visual tracking weight & \(K_{\text{vision}} = {1}/{n_{\text{tracked}}}\) \\
Preferred distance with neighbors & \(R_0 = 1/L_B\) \\
Social interaction torque coefficients & \(K_\textrm{AT} + K_\textrm{AL} = 30; K_{\text{wall}} = 10\) \\
The portion of attraction torque & \(a_T = {K_\text{AT}}/{(K_\text{AT} + K_\text{AL})}\) \\
PD control coefficients & \(K_p = 1, K_d = 3\) \\
Wall avoidance coefficient & \(K_w = 10\) \\
\toprule 
\end{tabular}
\end{adjustbox}
\label{table:parametersofwake}
\end{table}  
\renewcommand{\arraystretch}{1}
\begin{table}[ht]
\caption{\label{tab:table2}Parameters of the potential field of fish.}
\scriptsize 
\begin{adjustbox}{width=\textwidth,center}
\begin{tabular}{ll}
\toprule 
\textbf{Parameter} & \textbf{Expression} \\ \hline
Distance to the centroid & $a_1, a_2, a_3, a_4=0.40,0.10,0.30,0.475$ \\
Strength & $m_1, m_2, m_3, m_4=7.677(10^{-4}),7.726(10^{-4}),7.677(10^{-4}),4.562(10^{-4})$  \\
\toprule 
\label{table:parametersofpotential}
\end{tabular}
\end{adjustbox}
\end{table} 
\renewcommand{\arraystretch}{1}
\begin{table}[ht]
\caption{\label{tab:table3}Parameters of fish wake.}
\scriptsize 
\begin{adjustbox}{width=\textwidth,center}
\begin{tabular}{lll}
\toprule 
\textbf{Parameter} & \textbf{Expression} & \textbf{Comment}\\ \hline
Number of vortices per line & $N_\textrm{vor}=16$ & - \\
Generating time & $t_0$ & -  \\
Wake coefficient & $R_\textrm{wake}=0.005L_B$ & -  \\
Rankine vortex radius & $R_\textrm{vortex}=c_1L_B$ & $c_1=0.015$  \\
Spanwise velocity & \(V_s = c_1 V_\theta\) & \(c_1 = 0.25\) \\
Spanwise maximum & \(L_s = \pm V_s(t - t_0)e^{c_1(-t+t_0)}\) & \(c_1 = 0.1\) \\
Peak maximum & \(L_p = \pm \frac{c_1V_s}{t - t_0 + c_2}\) & \(c_1, c_2 = 4, 10\) \\
Swing direction & \(w_v = -1(\text{left}); 1(\text{right})\) & - \\
Vortex X & \(x_v \in [-L_s, L_s] \text{ uniformly by } N_{\text{vor}}\) & - \\
Vortex Y & $y_v = L_p + \frac{c_1w_v(\text{sin}^{c_2}(\frac{x_v}{L_s}\pi ))(t-t_0)}{t - t_0 + c_3} + \delta y_v$ & $c_1, c_2, c_3 = 0.3, 0.4, 10$ \\
Modification on vortex Y & $\delta y_v = c_1 + \frac{c_2\left|\frac{x_v}{L_s}\right|(t-t_0)}{t - t_0 + c_3}$ & $c_1, c_2 = -0.3, 0.6, 10$ \\
Initial circulation & $\mathit{\Gamma_0} = 2\pi V_s R_{\text{wake}}$ & - \\
circulation & $\mathit{\Gamma = \Gamma_0 e^{c_1(t-t_0-c_2)}}$ & $c_1, c_2 = -0.137, 0.4$ \\ 
Circulation distribution & $(\mathit{\Gamma_i} - -c_1)(t - t_0)/(t - t_0 + c_2)$ & $c_1, c_2 = 0.1, 0.5$\\
 & & Vortices at both ends are removed \\
Distance to wake vortices &$r_{\text{vor}}$& - \\
induced velocity & $u_w=\mathit{\Gamma}/(2\pi r_\textrm{vor})$ & - \\
\toprule 
\end{tabular}
\end{adjustbox}
\label{table:parametersoffishwake}
\end{table}  
\FloatBarrier
\quad
\bibliography{references}
\end{document}